# Plasmonics Theory for Biosensor Design: Mathematical Formulations and Practical Applications

Mariam Moussilli, Abdul Rahman El Falou, Raed Shubair

July 2018

# Contents

Abstract:

The last two decades have witnessed an exponential growth and tremendous developments in wireless technologies and systems, and their associated applications, such as those reported in [1]-[43]. In the recent years following 2006, there has been a great surge in interest in the newly emerging plasmonics nanotechnology because this new device technology provides tremendous synergy between electronic and photonic devices. Electronics devices are down-scalable up to the nanoscale size but have limited processor speed due to thermal and signal delay issues associated with electronic devices. On the other hand, photonic devices have extremely high speed and high data carrying capacity but are limited in size to the diffraction law such that the size of a photonic device should be equal to about half of its operational wavelength. The size mismatch between electronic devices and photonic devices inhibits the advantageous interfacing between these two device technologies and here plasmonics nanotechnology plays the important role of interfacing these two technologies. Plasmonics technology provides high speed interconnections with high data carrying capacity between nano-scale electronic devices opening a new field of research which is on-chip high speed nano-networks [28]. It is this great advantage of plasmonics technology that made it a very interesting technology for implementation for the design of a miniature real-time biosensor. In our plasmonic biosensor design, we utilized a subset of plasmonics technology which is surface plasmon wave generation in order to continuously monitor the concentration of a desired analyte.

In our project, we discussed the conditions for excitation of surface plasmon waves at a metal-dielectric interface and explained how to accurately find the frequency dependent dielectric function of the metal in the frequency ranges where the metal can support surface plasmon wave propagation. Then we explained the propagation constant of surface plasmon waves which is dependent on the dielectric constant of the metal and the dielectric constant of the dielectric medium at the interface where the surface plasmon waves propagate and we utilized this property of surface plasmon waves to excite these waves at the interface between the plasmonic active metal and a dielectric test sample to ultimately find the concentration of the analyte of interest from the dielectric constant of the test sample to which the analyte belongs. We explained the different coupling apparatuses and sensor probe configurations used for the excitation of surface plasmon waves and chose the most advantageous sensor probe and coupling apparatus format. After establishing the sensor probe and coupling apparatus format, we moved on to enhance the performance parameters of the selected biosensor format by the addition of a Graphene layer and by optimizing the sensor parameters. Then we adapted this sensor for continuous Glucose monitoring by implanting the disposable sensor probe into a blood vein of the diabetes patient and suggested instrumentation and safety measures to this biomedical application.

| Property 3: | **The dielectric function of SPWs depends of the dielectric function of the metal, the dielectric constant of the dielectric medium surrounding the metal and on the frequency of the incident light photons** |
|---|---|
| Property 4: | SPWs are purely longitudinal charge density oscillations at the metal-dielectric interface |

**Table 1- Summarization of the properties of SPWs**

As we can notice, the propagation constant of the SPWs at the metal-dielectric interface depends on the dielectric function of the metal and the dielectric medium surrounding the metal as well as on the frequency of the incident light photons. The dependence of the propagation constant of SPWs on the parameters mentioned will be utilized in order to design the plasmonic biosensor mechanism for finding analyte concentrations and will be discussed in detail later on in "2.3.4: Finding the Concentration of the desired Analyte".

Chapter 2: SPR Biosensors

In this chapter, we explain the advantages and general procedure followed by label-free biosensors in order to detect and monitor the concentration of the desired analyte in real-time and without compromising the constitution and properties of the sample.

Then we explain the conventional prism coupled SPR biosensors and how these sensors utilize the change in the resonance condition as a transduction mechanism to find the concentration of the desired analyte by angular integration and define the main performance parameters of the SPR sensors utilizing angle interrogation.

We then move on to explain FO-SPR biosensor advantages over prism coupling, construction, operation principle, transduction technique by spectral interrogation, performance parameters and finally explain how to find the concentration of a desired analyte within a sample by utilizing this FO-SPR biosensor.

## 2.1: Introduction to Biosensors

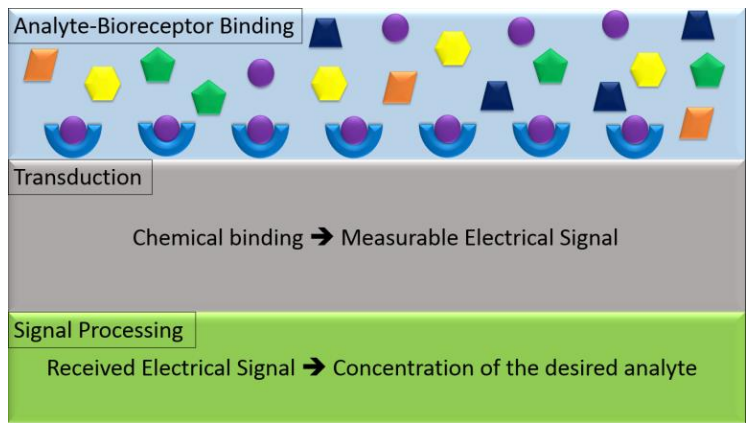

**Figure 1- The process of detecting a desired analyte and measuring its concentration by a label-free biosensor (top to bottom)**

Label-free biosensors detect the presence of the desired analyte and measure its concentration by studying physical optical, electrical or acoustic parameters while label based biosensors label the analyte of interest with labeling techniques such as fluorescence labeling or radiolabeling [40].

Due to the label-free technique used by label-free biosensors, the detection and monitoring of the concentration of the desired analyte is done in real-time and without compromising the sample constitution unlike label based biosensors [15]. The advantages of sample preservation and real-time analysis provided only by label-free sensing was the main motivation behind the surge in interest in designing highly optimized label-free sensors for biomedical applications.

Label-free biosensors are analytical devices based on the concept of transduction of the chemical interaction between a desired analyte and its bioreceptor into a measurable electrical signal in order to analyze and process this signal to find the concentration of the desired analyte [41].

The sample containing the analyte intended to be detected and measured is introduced to the label-free biosensor surface where bioreceptors that can only interact with the desired analyte are immobilized. Then a transduction mechanism changes the chemical binding of the desired analyte to its corresponding bioreceptor into a measurable electrical signal that could be analyzed and processed in order to find the concentration of the desired analyte [41].

The main interest in our study is to utilize the concept of SPWs of the Plasmonics nano-technology in order to design a highly accurate and highly sensitive label-free miniature nano-biosensor.

## 2.2: Conventional Prism Based Coupling

### 2.2.1: Necessity of Coupling Apparatuses

We shall assume that there is no coupling device and that the incident P-polarized light photons from the dielectric medium of Surrounding Refractive Index SRI and dielectric constant $\varepsilon_s$ surrounding the metal should excite SPWs at the metal-dielectric interface. For the excitation of the SPWs by the incident photons to occur, the incident photons should have the same momentum as the SPWs intended to be excited [42].

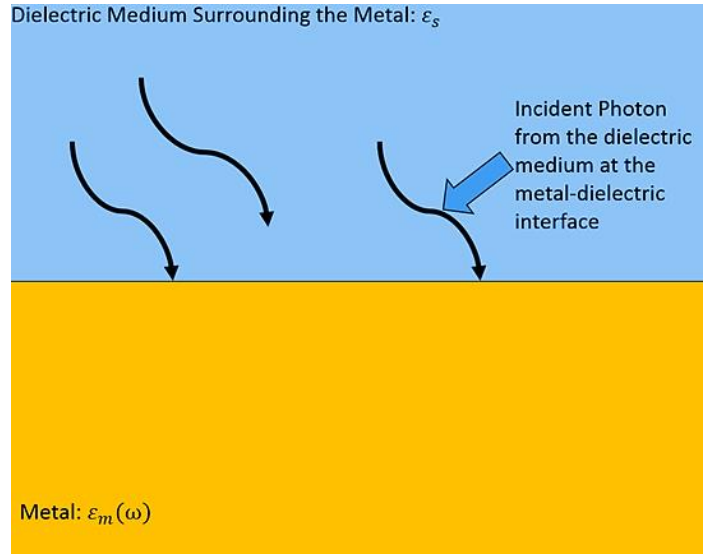

**Figure 2- the incident light photons from the dielectric medium at the metal-dielectric interface**

The momentum of a photon in the dielectric medium of refractive index SRI surrounding the Metal is:

$$p = \frac{h}{\lambda} = \frac{hf}{c} SRI = \frac{hf}{c}\sqrt{\varepsilon_s} = \hbar K_s \tag{11}$$

Where $h$ is the Planck's constant and $\hbar$ is the reduced Planck's constant.

The maximum propagation constant of the incident P-polarized light photon traveling through the dielectric medium is:

$$K_s = \frac{\omega}{c} SRI = \frac{\omega}{c}\sqrt{\varepsilon_s} \tag{12}$$

$\lambda$ is the wavelength of the incident light photons and $K_s$ is the propagation constant of the photon traveling through the dielectric medium of dielectric constant $\varepsilon_s$.

As shown in the photon momentum equation, the momentum $p$ is directly proportional to the propagation constant of the incident light photons incoming from the dielectric medium $K_s$ so in other words we can say that in order for the photons incident at the metal from the dielectric medium to excite SPWs at the metal-dielectric interface, the propagation constant of the incident light photons in the dielectric medium $K_s$ should be equal to the propagation constant of the SPWs $K_{SP}$ intended to be excited.

Form the conditions for the excitation of SPWs at the metal-dielectric interface, the real part of the dielectric function of the metal should be negative $Real[\varepsilon_m(\omega)] < 0$ and the real dielectric constant of the dielectric medium should be positive $\varepsilon_s > 0$ leading to the result that the propagation constant of the SPWs is greater than the propagation constant of the incident light photons from the dielectric medium when no coupling apparatus is used: $K_{SP} > K_s$.

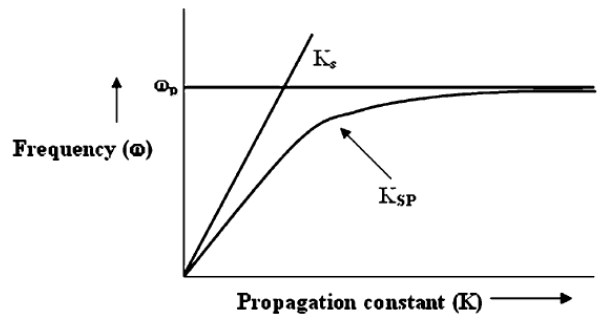

**Figure 3- plot showing the propagation constant of the SPWs to be excited at the metal-dielectric interface and the propagation constant of incident light photons as a function of frequency. This plot shows that the propagation constant of SPWs in always greater than the propagation constant of incident light photons without coupling so a coupling apparatus must be used in order to excite SPWs by increasing the momentum of the incident light photons to be equal to the momentum of the SPWs [8]**

Evidently without coupling, the P-polarized incident light photons incoming from the dielectric medium surrounding the metal cannot excite SPWs at the metal-dielectric interface so the use of a coupling device to increase the momentum of the incident light photons in order to excite SPWs is required and is critical for our biosensing application.

In this report we shall explain the conventional prism coupling and the more modern fiber optic coupling and analyze both coupling techniques in order to select the most advantageous coupling technique to implement in the design of our plasmonic biosensor.

## 2.2.2: Otto Configuration for the Prism coupled SPR Biosensor:

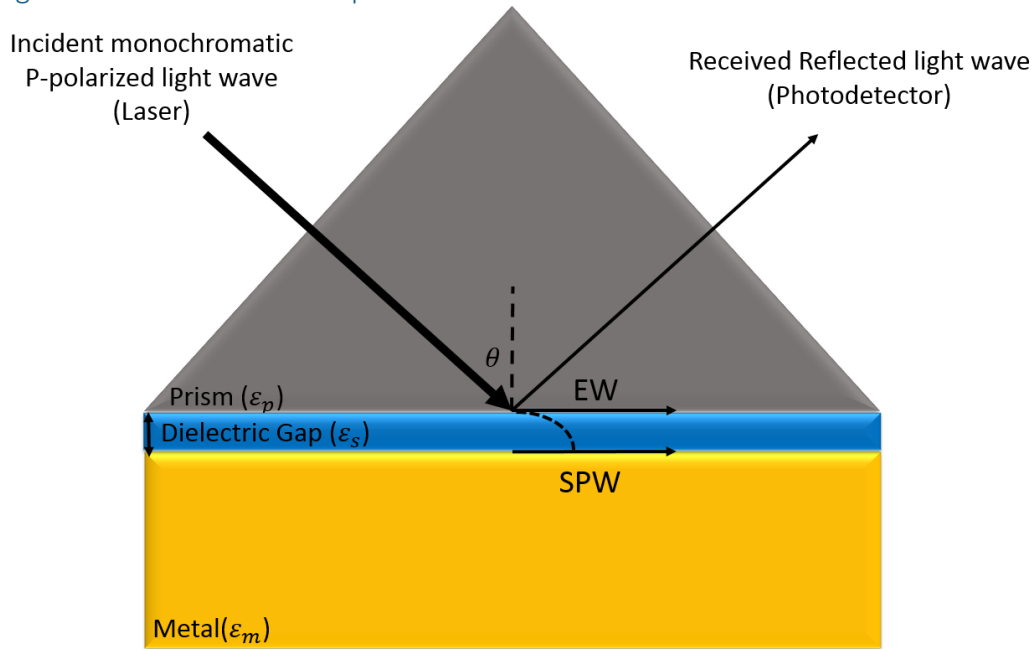

**Figure 4- The Otto prism based configuration of the SPR biosensor for excitation of SPWs at the metal-dielectric interface by EWs at the prism-dielectric interface**

The Otto configuration first proposed by [18] in 1968; is a prism based SPR biosensor configuration composed of a:

- Coherent mono-chromatic in-phase P-polarized laser light source
- Coupling bulk prism with high refractive index $n_p$ and a dielectric constant $\varepsilon_p$
- Dielectric medium gap between the base of the prism and the surface of the metal of lower refractive index $SRI$ compared to the prism and dielectric constant $\varepsilon_s$
- Surface plasmon active metal layer of dielectric function $\varepsilon_m(\omega)$
- Optical receiver and processing backhaul to analyze the reflected signal at the other side of the coupling prism

The Otto configuration is based on the concept of excitation of SPWs at the metal-dielectric interface by absorbed EWs at the prism-dielectric interface utilizing the concept Attenuated Total Internal Reflection (ATR) of the incident light waves at the prism-dielectric interface. The laser light enters through one side of the prism, is incident at the prism-dielectric interface at an angle greater than or equal to the critical angle for ATR to occur at the prism-dielectric interface and thus, part of the incident light at the prism-dielectric interface is absorbed as EWs and the other part is reflected back into the prism and exists the prism at the other side and is received by a an optical receiver for analysis and processing.

The absorbed EWs are bound to the prism-dielectric interface and decay exponentially in the prism as well as in the dielectric. EWs have the same properties as SPWs so, and there is a strong possibility of interaction between these two waves. The P-polarized component of the EWs is the component of EWs that is parallel to the plane of incidence and it is the only component of EWs that is able excites SPWs at the metal-dielectric interface.

In order for the P-polarized component of the EWs at the prism-dielectric interface to excite SPWs at the metal-dielectric interface, the dielectric gap between the base of the coupling prism and the surface of the metal should be: small enough to not significantly attenuate the EWs to the level that these waves cannot effectively excite SPWs by decaying notably in the dielectric gap before reaching the metal surface and should not be small to the level that is practically challenging to achieve.

The propagation constant of the P-polarized component of the EWs parallel to the plane of incidence [42]:

$$K_{EW} = \frac{\omega}{C}\sqrt{\varepsilon_p}\sin(\theta) \tag{13}$$

Where $\theta$ is the angle of incidence of the incident laser light at the prism-dielectric interface.

The Otto configuration has limited practicality since the dielectric gap needed is about 200 nm which is hard to physically realize and implement.

| Conditions for the excitation of SPWs by EWs utilizing the Otto configuration: | |
|---|---|
| **Condition 1:** | The coupling prism must have a refractive index greater than the refractive index of the dielectric medium gap in order for the coupling apparatus to increase the momentum of the incident light photons to reach the value of the momentum of the SPWs to be excited |
| **Condition 2:** | The dielectric gap between the base of the coupling prism and the surface of the metal should be small enough in order for the EWs to successfully excite SPWs at the metal-dielectric waves before decaying significantly while crossing the dielectric gap |
| **Condition 3:** | The dielectric gap should be decrease but not eliminated so the base of the prism and the surface of the metal should be in close proximity but not in contact |
| **Condition 4:** | The P-polarized component of the EWs at the metal surface after crossing the dielectric gap should have a propagation constant equal to the propagation constant of the SPWs in order for the EWs to excite SPWs |

Table 2- Conditions for the excitation of SPWs by EWs utilizing the Otto prism coupling configuration

## 2.2.3: Kretschmann-Reather Configuration of the Prism coupled SPR Biosensor

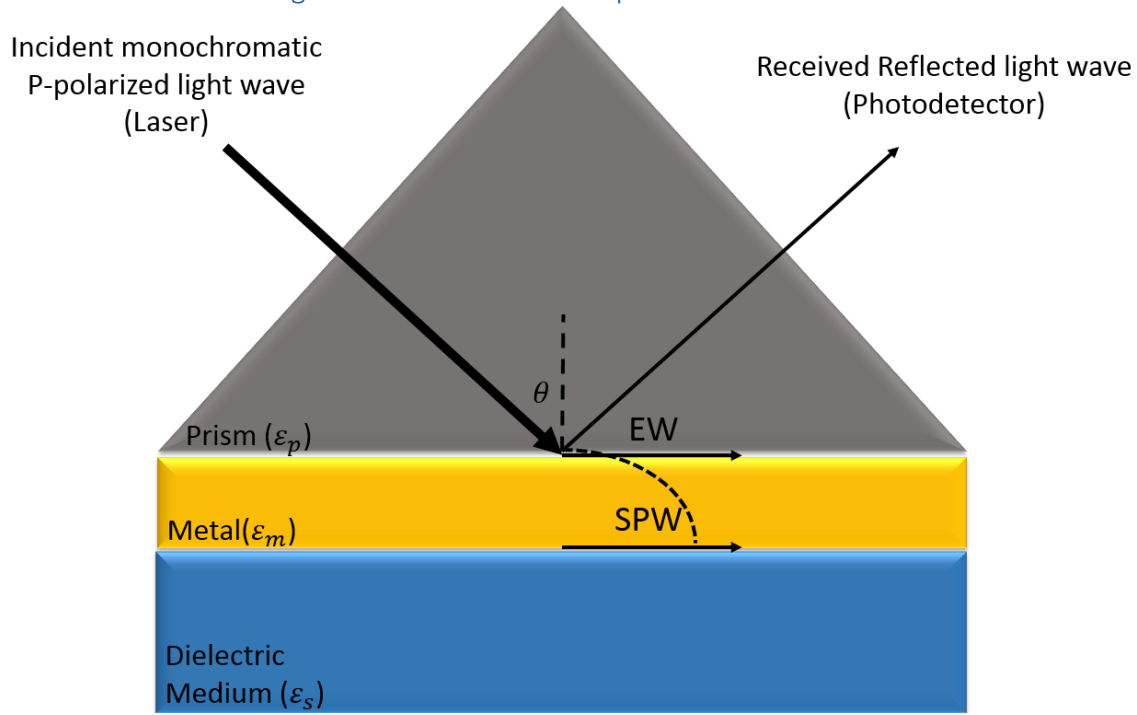

**Figure 5-The Kretschmann-Reather prism based configuration of the SPR biosensor**

The Kretschmann-Reather configuration first proposed by [44] in 1986; is based on the same concept as the Otto configuration which is the excitation of SPWs at the metal-dielectric interface by absorbed EWs but the Kretschmann configuration came as an improvement to the Otto configuration since this configuration eliminates the need for the unpractical and difficult to establish dielectric gap between the metal layer surface and the coupling prism base.

The Kretschmann-Reather configuration is composed of a:

- Coherent mono-chromatic in-phase P-polarized laser light source
- Coupling bulk prism with high refractive index $n_p$ and a dielectric constant $\varepsilon_p$
- Surface plasmon active metal layer of dielectric function $\varepsilon_m(\omega)$ directly coating the base of the coupling prism
- Dielectric medium of lower refractive index $SRI$ compared to the prism and a dielectric constant $\varepsilon_s$ surrounding the metal layer
- Optical receiver and processing backhaul for analysis of the reflected light wave off of the prism-metal interface

The dielectric medium is the dielectric sample introduced to the surface of the metal layer with the purpose of finding the concentration of an analyte of interest within this sample. The metal layer surface at the metal-dielectric interface is functionalized in order to have immobilized bioreceptors that only interact with the analyte of interest.

The incident light from the laser enters through one side of the coupling prism, is incident at the prism-metal interface at an angle of incidence equal to or greater than the critical angle for ATR to occurs then, part of the incident light wave is reflected to the other side the prism where an optical receiver is located to receiver and analyze the reflected light wave while the other part of the incident light wave is absorbed as EWs bound to the prism-metal interface and decay exponentially in the prism as well as in the metal.

The component of EWs parallel to the plane of incidence excites SPWs at the metal-dielectric interface given that the thickness of the metal is not thick enough (typically about 50 nm) to decay the EWs propagating through the metal to the level that these waves reach the metal-dielectric interface attenuated to a level that they cannot effectively excite SPWs at the metal-dielectric interface.

| **Conditions for the excitation of SPWs by EWs utilizing the Kretschmann-Reather configuration:** | |
|---|---|
| **Condition 1:** | The coupling prism must have a refractive index greater than the refractive index of the dielectric medium surrounding the metal from its surface that is not coated by the test sample (usually air) in order for the coupling apparatus to increase the momentum of the incident light photons to reach the value of the momentum of the SPWs to be excited |
| **Condition 2:** | The thickness of the metal should be small enough that it doesn't decay the EWs to the level that these EWs cannot excite SPWs upon reaching the metal-dielectric sample interface |
| **Condition 3:** | The metal should not be thin to the level that causes it to crack easily or prematurely because upon cracking the sensor lifetime will be compromised |
| **Condition 4:** | The P-polarized component of the EWs at the prism-metal interface should have a momentum equal to the SPWs intended to be excited at the metal-dielectric sample interface after the these EWs cross the metal to reach the metal-dielectric sample interface |

**Table 3- Conditions for excitation of SPWs at the metal-dielectric sample interface by EWs at the prism-metal interface**

## 2.2.4: Resonance Condition and Angular Interrogation

As mentioned above in "2.2.1: Necessity of Coupling Apparatuses", in order for the excitation of SPWs to occur then the P-polarized EWs should have the same propagation constant as the SPWs intended to be excited. For the case of prism coupling, the component of the absorbed EWs parallel to the plane of incidence should have the same propagation constant as the SPWs to be excited at the metal-dielectric interface.

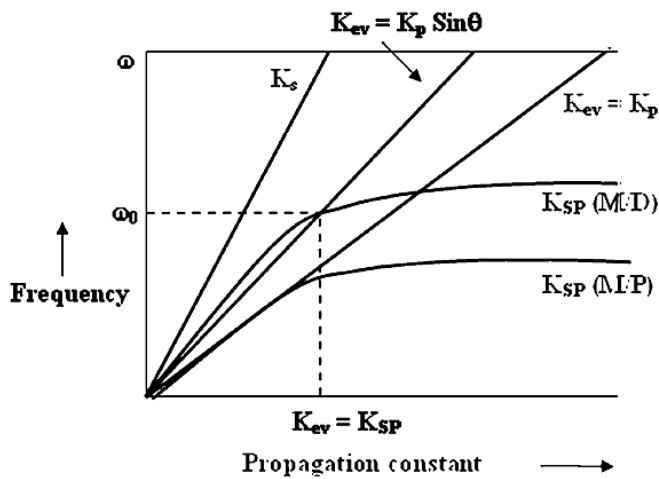

**Figure 6-the condition for excitation of SPWs is that the propagation constant of the SPWs must be equal to the EWs propagation constant component that is parallel to the plane of incidence which disables the prism-metal interface (M/P) from supporting SPWs and disables incident photons without coupling from exciting SPWs at any interface [8]**

From "Figure 6" and only considering the Kretschmann-Reather configuration, we can see that the uncoupled photons incident at the metal surface from the dielectric medium surrounding the metal cannot excite SPWs at neither the metal-dielectric interface nor the metal-prism interface because the propagation constants for the SPWs at both the metal-prism interface and the metal-dielectric interface are greater than the propagation constant of the uncoupled incident light photons incoming from the surrounding dielectric medium.

Meanwhile for prism coupling, the P-polarized component of the EWs for angles of incidence ranging from the critical angle for ATR to occur at the prism-metal interface to 90° can only excite SPWs at the metal-dielectric interface since:

- The propagation constant of the EW component parallel to the plane of incidence is equal to the propagation constant of the SPWs at the metal-dielectric interface for light of angular frequency $\omega_0$ and incidence angle $\theta$
- The propagation constant of the P-polarized component of the EWs at the prism-metal interface is always less than the propagation constant of SPWs at the prism-metal interface so no SPWs are excited at the prism-metal interface

The condition for excitation of SPWs by the absorbed and exponentially decaying EWs of same propagation constant is known as the resonance condition and leads to maximum coupling of the EWs to the SPWs. The resonance condition for prism based coupling is:

$$\frac{\omega}{C}\sqrt{\varepsilon_p}\sin(\theta_{res}) = \frac{\omega}{C}\sqrt{\frac{\varepsilon_m(\omega) \times \varepsilon_s}{\varepsilon_m(\omega) + \varepsilon_s}} \quad (14)$$

Where $\theta_{res}$ is the incidence angle $\theta$ when the resonance condition is satisfied.

In the prism based design the light source is a mono-chromatic laser so, the angular frequency of the incident light photons $\omega$ is a fixed value depending on the wavelength of the used laser. The dielectric function of the metal is frequency dependent but since the laser light source in mono-chromatic then the value of the dielectric function of the metal is fixed to a value corresponding to the wavelength of the laser used. So, the parameters varying the resonance condition are: the angle of incidence $\theta$ and the dielectric constant of the dielectric sample medium surrounding the metal layer.

The resonance condition is used to detect the variation in the refractive index of the medium surrounding the metal layer:

$$SRI = \sqrt{\varepsilon_s} \quad (15)$$

When the SRI of the dielectric sample medium changes, the incidence angle at which the resonance condition is satisfied $\theta_{res}$ and maximum excitation of SPWs by EWs occurs is altered and thus the incidence angle is varied until the resonance condition is re-established. When the resonance condition is re-established, the new value of the SRI of the dielectric sample is found by the resonance condition equation.

The method of changing the angle of incidence until the resonance condition is re-established is called angular interrogation and is used primarily in prism based SPR biosensors to detect the variation of the refractive index of the dielectric sample medium introduced to the surface of the metal layer.

In order to utilize the resonance condition under the angular interrogation technique to find the concentration of a desired analyte, bioreceptors specific to this analyte are immobilized at the metal surface so that the SRI of the dielectric sample only varies when the interaction of the desired analyte with its bioreceptors varies. Then after finding the new refractive index of the dielectric sample there is a unique mapping between the refractive index of a sample and the concentration of the desired analyte within this sample [45,46].The mapping differs for different dielectric samples, bioreceptors and analytes of interest.

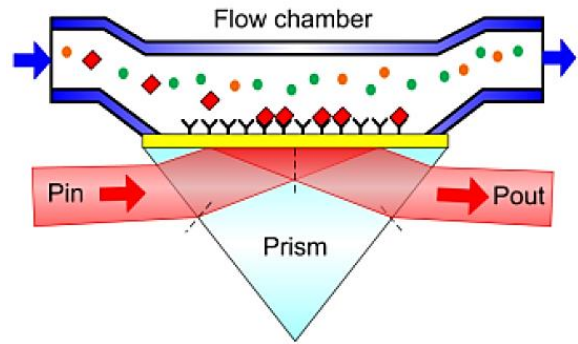

**Figure 7- the Kretschmann-Reather prism coupled SPR biosensor with a bioreceptors immobilized on the surface of the metal layer in contact with the dielectric sample in order to vary the refractive index of the sample only according to variations in the interaction between the bioreceptors and their analyte of interest [47]**

## 2.2.5: Response Curve for Prism Based SPR Biosensor for Angular Interrogation

Considering the more widely used Kretschmann-Reather prism based SPR biosensor configuration:

At the resonance angle of incidence $\theta_{res}$, the resonance condition is satisfied so there is maximal excitation of SPWs at the metal-dielectric sample interface by EWs bound to the prism-metal interface and exponentially decaying in the metal. Since EWs are absorbed waves from the incident light due to ATR of the incident light photons at the prism-metal interface hence; at the resonance condition there is maximal coupling of the incident light photons at the prism-metal interface to the SPWs at the metal-dielectric interface.

The only output signal that is available for processing and analysis is the light wave reflected off of the prism-metal interface due to ATR of the incident light wave at the prism-metal interface. So, the angle of incidence at which the resonance condition occurs must be deduced from the reflected light wave or typically the normalized reflected light wave.

When the incident light wave is incident on the prism-metal layer by an angle equal to or greater than the critical angle for ATR, ATR occurs and part of the incident light wave is reflected back into the prism and exits the prism to be received by the analysis and processing block of the biosensor. The other part of the incident light wave is absorbed as an EW and traverses the prism-metal interface. So, due to ATR we can say that the incident light wave power is divided to a reflected part of normalized reflected power $R$ and an absorbed part with normalized absorbed power $A$ [33]:

$$R + A = 1 \quad\quad\quad\quad (16)$$

Building on the conclusions drawn above, at the resonance condition there will be maximal absorption $A$ and minimal reflectance $R$ of the incident light photons at an incidence angle $\theta_{res}$ between the critical angle for ATR and 90°.

So, when plotting the response curve of the sensor which is the normalized received power with respect to the incidence angle there will be a sharp dip in the received power $R$ at the resonance angle $\theta_{res}$ [42] as shown below in "Figure 8".

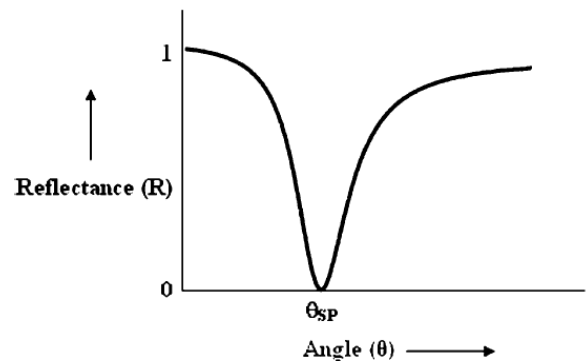

**Figure 8- The response curve of the prism based SPR sensor showing a dip in the reflected normalized power at the resonance angle [42]**

## 2.2.6: Performance Parameters for Prism based SPR Biosensors Utilizing Angular Interrogation

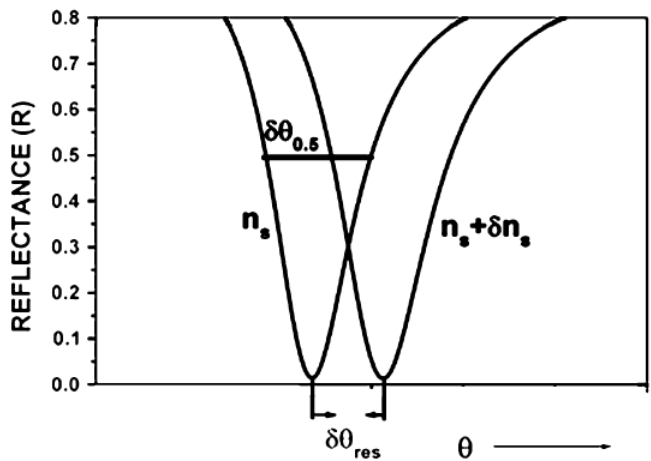

**Figure 9- Variation in the SRI of the dielectric sample causing a red-shift in the reflectance dip and broadening of the angular width due to a shift in the resonance angle at which the resonance condition is satisfied [33]**

*Sensitivity of the prism coupled SPR biosensor based on angular interrogation:*

When the propagation constant of the SPWs to be excited is equal to the propagation constant of the component of EWs that is parallel to the plane of incidence, then there is maximum excitation of SPWs by the component of EWs parallel to the plane of incidence and thus, the resonance condition is satisfied for a particular angle of resonance $\theta_{res}$.

The propagation constant of SPWs is very sensitive towards variations in the SRI of the dielectric sample medium surrounding the metal so when the SRI of the dielectric sample varies, it causes the propagation constant of the SPWs to vary as well leading to a shift in the resonance angle at which the resonance condition is satisfied.

The sensitivity $S_{n/degree}$ (measured in degrees/RIU) of the prism coupled SPR biosensor based on angular interrogation is the shift in the resonance angle $\delta\theta_{res}$ due to a variation in the SRI $\delta SRI$ of the dielectric sample surrounding the metal [8] [17]:

$$S_{n/degree} = \frac{\delta\theta_{res}}{\delta SRI} \quad (17)$$

This change in the SRI of the dielectric sample is detected by the shift in the resonance angle at which the dip in the normalized reflected power $R$ occurs.

The greater the sensitivity, the better the sensor is at detecting small variations in the SRI of the dielectric sample because sensors with high sensitivity translate minor variations in the SRI of the dielectric sample into pronounced shifts in the resonance angle.

*Detection Accuracy of the prism coupled SPR biosensor based on angular interrogation:*

The detection accuracy $DA_{/degree}$ of the prism coupled SPR biosensor based on angle interrogation is the shift in the resonance angle $\delta\theta_{res}$ with respect to the angular width of the response curve of the biosensor before shifting at 50% $\delta\theta_{0.5}$ reflectance $R$ [33,42]:

$$DA_{/degree} = \frac{\delta\theta_{res}}{\delta\theta_{0.5}} \qquad (18)$$

When the SRI of the dielectric sample surrounding the metal varies, the dip in reflectance of the response curve of the sensor is shifted due to the shift in the resonance angle at which the resonance condition is satisfied. The higher the $DA$ of the sensor, the more accurate and precise the sensor is at locating the shift in the resonance angle in order to find the variation in the SRI of the dielectric sample and to map it to find an accurate value for the concentration of the desired analyte.

2.3: Fiber Optic Based Coupling

2.3.1: The Advantages the FO- SPR Biosensors

Light waves incident at the fiber core with angles of incidence belonging to the acceptance cone of the optical fiber, are guided by the optical fiber. The main purpose of optical fibers is to utilize the Total Internal Reflection TIR principle to guide the accepted light waves through the core of the optical fiber by reflecting these waves off of the cladding and back into the fiber core until the other end of the optical fibers is reached [48].

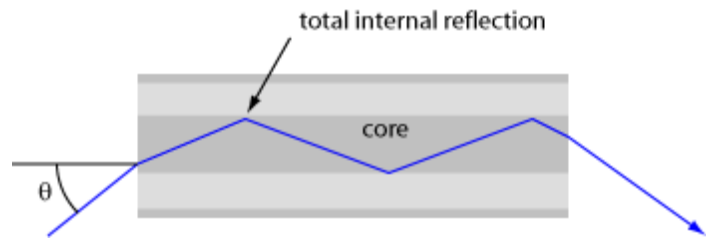

**Figure 10- the optical fiber guiding the light wave by total internal reflection at the core cladding interface where the light wave incident at the core-cladding interface is reflected off of the cladding and back into the core [23]**

Considering the Kretschmann-Reather configuration, the main purpose of the coupling prism was to cause ATR of the incident light waves at the prism-metal interface in order to generate EWs at the prism-metal interface by absorption of part of the incident light wave and to reflect the other part of the incident light wave to the processing block of the biosensor for analysis. Hence as first suggested by [49] in 1993, the core of the optical fiber can serve as an advantageous and logical replacement for the coupling prism used in conventional SPR sensors.

The use of an optical fiber core instead of a prism as the coupling apparatus has many advantages such as [49]:

- Significantly decreasing the weight of the biosensor due to the replacement of the bulky heavy weight coupling prism by an extremely light weight fiber core increasing the practicality of the sensor particularly for in situ measurements
- Drastically decreasing the dimensions of the biosensor to become miniature for easy for installation and increase in adaptability
- Facilitating the possibility for remote sensing
- Enabling the use of smaller sample sizes which is particularly important in biomedical applications
- Effectively decreasing the complicity of the processing backhaul
- Decreasing the complexity of the optical design of the sensor
- Opening up the possibility for adaptation of the sensor for biomedical applications that require the sensor to be disposable, implantable, nano-sized and so on …
- Increasing the flexibility of the design of the biosensor

Due to these numerous advantages, we decided to adopt the optical fiber core as the coupling apparatus for the design of our SPR biosensor. The design, optimization and adaptation of our proposed SPR biosensor will be discussed in detail throughout the remainder of the chapter.

## 2.3.2: Conventional Design of the FO- SPR Biosensor

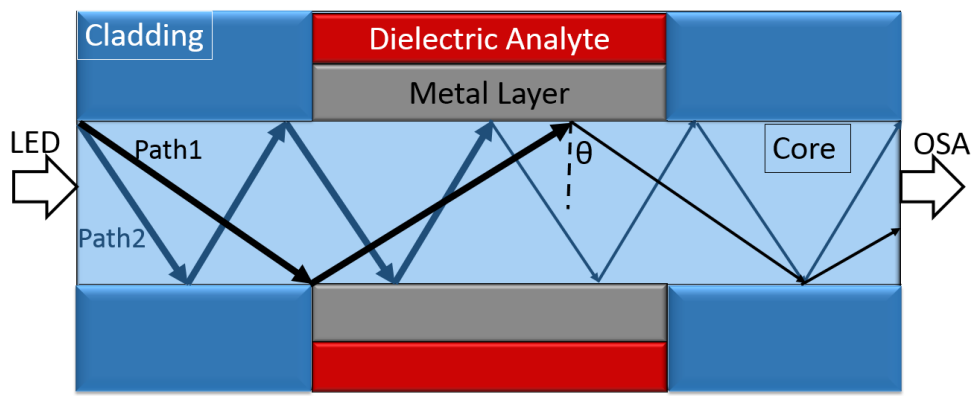

**Figure 11- the conventional design of a FO- SPR biosensor**

The conventional design of the FO- SPR biosensor is composed of: a glass Multi-Mode optical Fiber (MMF) as the coupling apparatus and remote sensing facilitator, a poly-chromatic Led as the light source, a metal layer (usually Gold), a dielectric sample and an Optical Spectrum Analyzer (OSA) as the processing backhaul of the biosensor [50].

Despite the existence of multiple FO-SPR biosensor designs utilizing single mode fibers such as in [51], we will focus on the MMF based conventional design in explaining the theory behind FO coupled SPR biosensors because MMFs facilitate easier coupling between fibers and thus, serve our goal of designing a disposable FO coupled SPR biosensor.

In order to form the sensing region of the biosensor, a part of the optical fiber cladding (usually made from Silica) is removed usually at the middle of the MMF then the exposed un-cladded core is covered by a metal layer. The metal layer is then exposed to the dielectric sample for sensing the concentration of the analyte of interest in this sample. The portions of the optical fibers going from and leading to the sensing region are used to facilitate remote sensing by guiding the light wave by TIR to reach the desired destination [33].

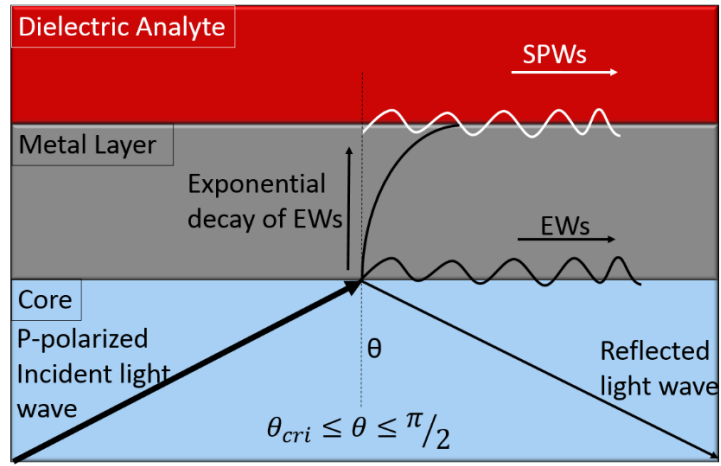

**Figure 12-the excitation of SPWs at the metal-dielectric interface by EWs at the core-metal interface**

The operation principle of the FO coupled SPR sensor is shown in the figure above and is the same as the principle of operation of the Kretschmann-Reather configuration based prism coupled SPR biosensor [45]:

The optical fiber only guides the light waves that are:

- incident at the optical fiber core with an angle within the acceptance cone of the fiber noting that the acceptance cone of the fiber depends on the characteristic numerical aperture of the fiber that is in itself a function of the values of the refractive indexes of the core and cladding
- incident at the core-cladding interface with an angle of incidence $\theta$ between the critical angle for TIR to occur at the core-cladding interface $\theta_{cri}$ and 90°

When the guided light waves reach the sensing layer, ATR occurs at the core-metal interface. ATR of the incident light waves at the core-metal interface causes:

- part of the incident light waves to be absorbed as EWs bound to the core-metal interface and decay in the core as well as in the metal
- other part of the incident light waves to be reflected back into the optical fiber core to be guided by the cladded optical fiber core to reach the OSA for analysis and plotting

The EWs at the core-metal interface excite SPWs at the metal-dielectric interface as long as the thickness of the metal layer is not thick enough to decay the EWs to the level that they cannot excite SPWs when reaching the metal-dielectric interface [52].

At the resonance condition, surface plasmon excitation happens when the energy and momentum conservation of the quantum particles is observed when the photons are absorbed and transformed into plasmons [39].

The resonance condition for maximal excitation of SPWs at the metal-dielectric interface by the component of the absorbed EWs parallel to plane of incidence at the core-metal interface is that the propagation constant of the component of exciting EWs parallel to the plane of incidence $K_{EvaW}$ should be equal to the propagation constant $K_{SP}$ of the SPWs [39]:

$$\frac{2\pi}{\lambda} \times n_{co} \times \sin(\theta) = \frac{2\pi}{\lambda} \sqrt{\frac{\varepsilon_m(\omega) \times \varepsilon_s}{\varepsilon_m(\omega) + \varepsilon_s}} \tag{19}$$

Where $n_{co}$ is the refractive index of the optical fiber core, $\lambda$ is the wavelength of the incident light in vacuum and $\theta$ is the angle of incidence of the incident light waves at the core-metal interface and ranges from $\theta_{cri}$ to 90°.

When the resonance condition is satisfied there will be maximum absorption of incident light waves at the core-metal interface in the form of EWs bound to the core-metal interface and there will be minimal reflectance detected by the OSA.

The interrogation method used by FO-SPR biosensors to detect any variation in the $SRI$ of the dielectric sample surrounding the metal is the spectral interrogation method also known as the multiple wavelength regime [52,53]. FO-SPR biosensors usually utilize a MMF core as the coupling apparatus hence incident light waves of multiple modes (paths) and angles of incidence ranging from $\theta_{cri}$ to 90° are supported and guided within the MMF core. So because of the utilization of a MMF, there will be incident light waves of incidence angles ranging from $\theta_{cri}$ to 90° incident at the sensing region of the biosensor so, there is no definite angle that can be pinpointed and determined as the angle at which resonance occurs opposite to prism coupled SPR biosensors utilizing mono-chromatic lasers of one mode (one path) as light sources [8]. FO-SPR biosensors also utilize a polychromatic light Led source so, there will also be light waves of different wavelengths incident at the sensing region [33]. Since OSAs can serve as a very helpful tool in displaying the normalized reflected light power for each wavelength in the received signal after incidence with the sensing metal layer; spectral interrogation method is used to pinpoint the resonance wavelength $\lambda_{res}$ at which the resonance condition is satisfied.

At the resonance wavelength the OSA displays a sharp dip in the normalized power of the received light waves after ATR at the core-metal interface. This is because the light waves that are incident at the core-metal interface with a wavelength equal to the resonance wavelength will be maximally absorbed as EWs at the core-metal to maximally excite SPWs at the metal-dielectric interface [52].

The propagation constant of SPWs at the metal-dielectric interface is very sensitive to any change in the $SRI$ of the dielectric sample surrounding the metal hence for any change in the $SRI$ of the metal, there will be shift a shift in the resonance wavelength $\lambda_{res}$ at which maximal absorption in the form of EWs and minimal reflectance detected by the OSA take place [52].

In order for the FO-SPR biosensor to detect the concentration of a specific analyte of interest within the dielectric sample surrounding the metal layer, bioreceptors specific only to the analyte of interest are immobilized at the metallic surface of the sensor probe. Due to the use of the bioreceptors that can only interact with the analyte of interest within the sample, then only an increase in the concentration of the desired analyte will induce an increase in the $SRI$ of the dielectric sample surrounding the metal layer and thus a red-shift in the resonance wavelength [54].

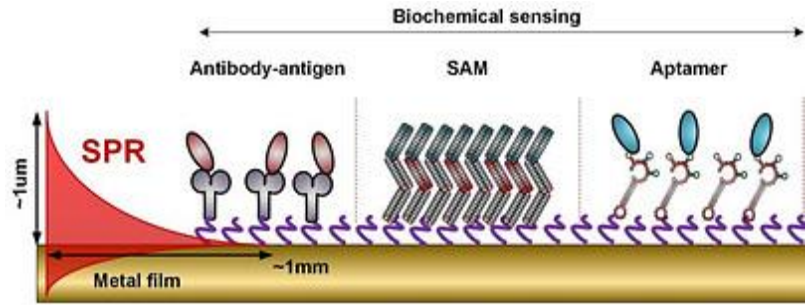

Figure 13- Examples of bioreceptors that could be immobilized at the metal surface in contact with the dielectric sample in order to interaction only with the analyte of interest in the sample [55]

The sensor probes the interaction of the desired analyte with its bioreceptors by the variation in the propagation constant of the SPWs and thus the variation in the resonance condition. The sensor is used as a transduction mechanism to translate the variation in the concentration of the desired analyte that imposes a variation in the refractive index of the sample into a variation in the resonance wavelength. When the resonance wavelength is detected, the resonance condition is used to find the corresponding refractive index of the sample then unique mapping depending on the analyte of interest and bioreceptor used is performed to find the concentration of the desired analyte [54].

2.3.3: Performance Parameters for the FO Coupled SPR Biosensor Utilizing Spectral Interrogation

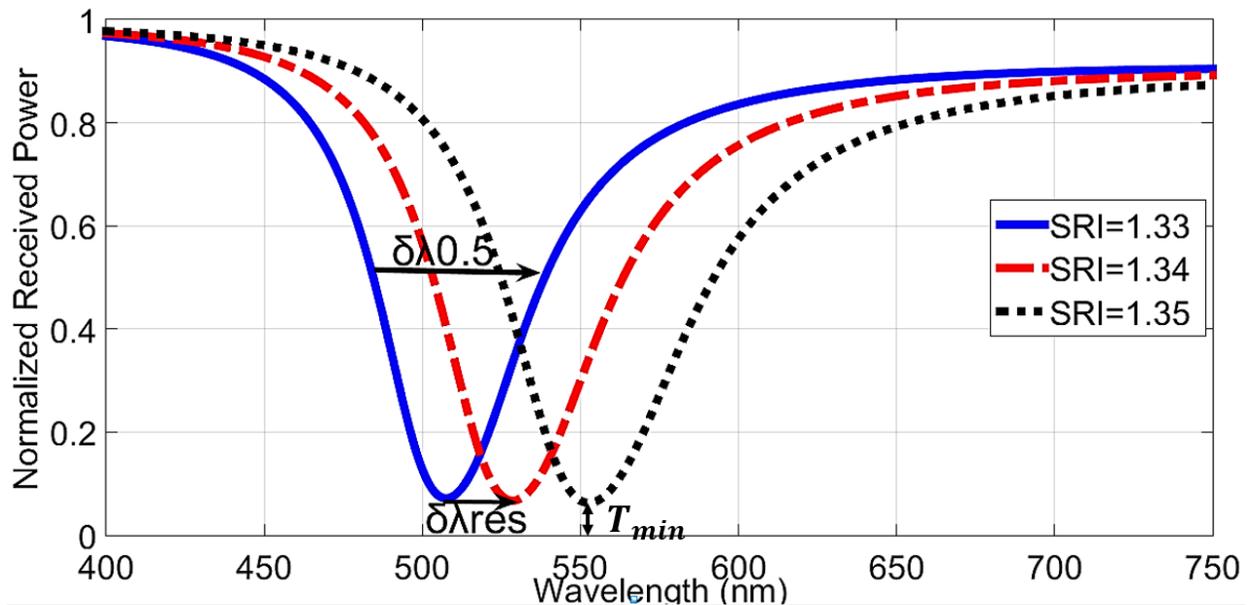

Figure 14- Red-shift in the resonance wavelength towards the higher wavelengths, broadening of the spectral width of the response curve of the sensor and increase in the depth of the dip of the response curve as the SRI of the dielectric sample increases

*Sensitivity of the sensor using spectral interrogation:*

In our design of an FO-SPR biosensor, we shall utilize a MMF core as the coupling apparatus since MMFs are easier to couple and the characteristic of easy coupling between MFFs will be used to facilitate an easier and more practical design for a disposable sensor probe.

Since a MMF core will be used as the coupling apparatus, then all light waves that are incident at the fiber core within the acceptance cone of the MMF will enter to the fiber core and all light waves that entered into the fiber core and have an angle of incidence between the critical angle for TIR at the core-cladding interface and 90° will be guided and supported by the MMF. So, the incident light waves at the core-metal interface will have angles ranging from $\theta_{cri}$ to 90°. Hence angular interrogation is not possible because we cannot pinpoint the exact mode (path) angle of incidence that caused resonance and dip in the reflectance. Since the OSA provides plotting of the received reflected normalized power off of the core-metal interface for each wavelength provided by the poly-chromatic Led light source so, we can precisely know the wavelength of the incident light photons that are maximally absorbed to excite surface plasmons.

Building on the analysis stated above, the shift in the resonance wavelength $\delta\lambda_{res}$ caused by the variation in the refractive index of the dielectric sample surrounding the metal $\delta SRI$ will be used to calculation of the sensitivity $S_n$ of the FO COUPLED SPR biosensor [52]:

$$S_n = \frac{\delta\lambda_{res}}{\delta SRI} \tag{20}$$

The greater the shift in the resonance wavelength as a response to the variation of the refractive index of the dielectric sample surrounding the metal, the better the sensor is at detection of even small variations in the refractive index of the dielectric sample and thus small variations in the concentration of the desired analyte.

*Detection Accuracy of the sensor using spectral interrogation:*

The detection accuracy of the sensor is the performance parameter of the sensor that determines how accurate and precise the sensor is at pinpointing the exact value of the resonance wavelength where the dip in reflectance occurs [47].

In order to find the most accurate estimation of the resonance wavelength corresponding to the dip of reflectance in the response curve of the sensor at a particular $SRI$ of the dielectric medium, the spectral width of the response curve should be as narrow as possible and the depth of the dip should be as deep as possible [47].

In order to take the conditions of accurate detection of the resonance wavelength into consideration, the $DA$ of the sensor is [47]:

$$DA = \frac{100 - T_{min}}{\delta\lambda_{0.5}} \tag{21}$$

Where $T_{min}$ is the percentage of minimum normalized received power at the resonance wavelength displayed by the OSA and $\delta\lambda_{0.5}$ is the spectral width of the response curve of the sensor at 50% normalized received power.

The greater the detection accuracy of the sensor, the better the sensor is at accurately estimating the resonance wavelength to find the best estimate of the refractive index of the sample and thus the best estimate of the concentration of the desired analyte in this sample.

*Resolution of the sensor using spectral interrogation:*
The resolution of the sensor is a very crucial performance parameter because it detects the minimum quantity of variation that a sensor is able to detect.

When considering our plasmonic sensor, the main purpose of the sensor is to measure the variation in the refractive index $\delta SRI$ of the dielectric sample surrounding the surface plasmon active metal. The plasmonic sensor does this by transducing the change in the refractive index of the sample $\delta SRI$ to a shift in resonance wavelength $\delta\lambda_{res}$ at which minimum received power is detected by the OSA.

Thus the greater the shift in resonance wavelength, the greater the number of sampling points taken by the OSA and thus smaller variations in the $SRI$ could be detected by the sensor.

Building on the above analysis, the resolution $R_{SRI}$ of the plasmonic biosensor is [25]:

$$R_n = \frac{\delta SRI}{N_{total}} = \frac{\delta SRI}{\delta\lambda_{res} \times N_{/1nm}} \qquad (22)$$

Where $N_{total}$ is the total number of sampling points taken by the OSA for the shift in the resonance wavelength $\delta\lambda_{res}$ induced by the change in the refractive index of the metal $\delta SRI$.

The total number of sampling points taken by the OSA for the shift in the resonance wavelength $\delta\lambda_{res}$ is the shift in the resonance wavelength multiplied by the number of sampling points taken by the OSA for 1 nm variation in the resonance wavelength $N_{/1nm}$.

The higher the SRI resolution, the lower the amount of variation of the SRI that could be detected by the sensor.

## 2.3.4: Finding the Concentration of the desired Analyte

For spectral interrogation, the FO-SPR biosensor transduces the change in the concentration of the desired analyte into a variation in the resonance wavelength at which minimum reflected power is observed by the OSA.

Upon change in the concentration of the analyte of interest, the interaction between the analyte of interest and its corresponding immobilized bioreceptors changes causing a changing in the refractive index of the sample surrounding the surface plasmon active metal.

Upon the variation of the refractive index of the dielectric sample surrounding the metal, the propagation constant of SPWs intended to be excited at the metal-dielectric sample interface varies. So, the resonance condition is varied and thus the resonance wavelength of the EWs that excite SPWs is altered.

When ATR occurs at the core-metal interface part of the incident light waves is absorbed as EWs bound to the core-metal interface and exponentially decay in the metal as well as in the dielectric and the other part of the incident light waves is guided by the fiber until it is received at the OSA. The absorbed light waves excite SPWs at the metal-dielectric interface and maximum absorption of the incident light waves occurs for the light waves with wavelength equal to the resonance wavelength. Hence the reflected light wave that is plotted by the OSA displays minimal received power at the resonance wavelength because the component of the light wave at the resonance wavelength was absorbed to couple to the SPWs.

In the table below, we summarize the procedure followed to find the concentration of the desired analyte from the resonance wavelength [29] [27]:

| **Steps to finding the concentration of the desired analyte utilizing the FO-SPR biosensor:** | |
|---|---|
| **Step 1: Determination of the resonance wavelength** | The only output signal available for analysis is the reflected light wave off of the core-metal interface due to ATR at the core-metal interface.<br>This reflected light wave is transmitted through the optical fiber until it is received by the OSA. The OSA then provides a plot of the normalized received power at each wavelength. The resonance wavelength is the wavelength at which minimal power is received. The accuracy in determining the exact wavelength at which minimal reflected power is received depends on the resolution of the OSA. |
| **Step 2: Determination of the refractive index of the sample** | By determining the resonance wavelength, the resonance condition is applied in order to find the refractive index of the dielectric sample surrounding the metal. |
| **Step 3: Determination of the Concentration of the desired analyte** | Upon determining the refractive index of the dielectric sample surrounding the metal, a specific mapping is applied to find the concentration of the desired analyte within a sample depending on the sample characteristics before introducing any analyte |

| | concentration and depending on the characteristics of the bioreceptors used to interact with the analyte of interest. |
|---|---|

**Table 4- summarization of the procedure used to find the concentration of the analyte of interest by finding the resonance wavelength induced by the change in the refractive index of the dielectric sample surrounding the metal**

Chapter 3: Graphene for the Enhancement of the Sensitivity of the FO-SPR Biosensor

In this chapter we state some of the main properties of Graphene, discuss recent studies on the addition of Graphene to FO-SPR biosensors, explain the constitution of the Graphene enhanced and FO-SPR sensor model, and propose an analysis model for finding the normalized reflected power due to ATR at the core-metal interface for Graphene enhanced and FO-SPR biosensors.

Then we study the response curves of the FO-SPR biosensor before and after addition of Graphene for both the addition of the Graphene layer for both the Gold and the Silver metal layers.

Finally we decide which metal to use for the optimization of our sensor and decide if we should add a Graphene protecting and absorbing layer.

## 3.1: Properties of Graphene

Graphene is a 0.34 nm one atom thick layer of Graphite in the form of a 2-deminsoinal (from the view of the electron) atomic-scale honeycomb hexagonal lattice constituted from stable carbon atoms [31].

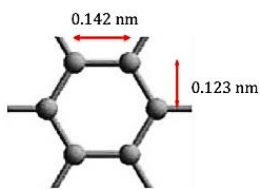

**Figure 15-the dimensions of the honeycomb crystal structure of Graphene sheets [31]**

Graphene has many extraordinary and advantageous properties that we summarize as:

- High Strength and flexibility:

Each carbon atom has 4 valence electronics where 1 valence electron of each carbon atom forms a single covalent bond with the neighboring carbon atoms' valence electrons and since covalent bonds are the strongest bonding type; graphene is thus very strong (stronger than diamonds) and has high elastic deformation since covalent bonds are very difficult to break and thus, can withstand allot of mechanical strain [57].

- High Electrical and Thermal Conductivity:

The free electrons traveling on the honeycomb crystal lattice of Graphene are traveling theoretically in the speed of light due to the loss of the effective mass of these quasiparticles leading to high electrical and thermal conductivity in Graphene (conducts electricity better than Silver and conducts heat better than diamonds) and high electron-hole mobility (higher than silicon) [57].

- Extremely Lightweight:

Graphene has a theoretical specific surface area volume of 2630 m2/g meaning that it is extremely lightweight [32].

- Low susceptibility to corrosion and high absorption:

Graphene is highly permeable making it withstand many environmental factors (better than Silver that oxidizes and corrodes rapidly) and transparent but has a very high absorption (higher than Gold) relative to its opacity and thickness [57].

- High Biocompatibility:

Graphene is biocompatible with the human body meaning that it does not produce a toxic or immunological response when exposed to the body or bodily fluids [57].

- High functionalization:

Graphene has a high ability to be functionalized which means that it possesses the ability of adding new functions, features, capabilities, or properties to a material by changing the surface chemistry of the material. It should be noted that both graphene and graphene oxide (GO) are versatile materials for functionalization [57].

The remarkable electronic and photonic properties of graphene have made it a material of great interest for a wide range of applications and in our application we shall focus on how Graphene can functionalize the metal layer of the FO-SPR sensor to improve the sensitivity and lifetime of this sensor.

## 3.2: Introduction of a Graphene layer to the FO-SPR Biosensor

The typical metals used for the design of SPR biosensors are: Gold and Silver. Gold is immune to corrosion and oxidation but has low absorption of biomolecules due to its chemical stability thus decreasing the sensitivity of the Gold based SPR biosensor [54]. Also Gold has a large absorption coefficient leading to the broadening of the dip in reflectance causing a decrease in the detection accuracy of the sensor [58]. On the other hand, Silver exhibits higher absorption than Gold thus increasing the sensitivity of the SPR sensor but has the disadvantage of being susceptible to oxidation and corrosion as time passes thus, decreasing the lifetime of the sensor and causing the plasmonic signal to vary with time due to variation in the Silver metal dielectric function due to corrosion and oxidation [58]. Also Silver has a lower absorption coefficient leading a narrower response curve and higher detection accuracy compared to Gold based sensors [58].

Concerning FO-SPR biosensors utilizing a Gold type metal layer, recent studies such as [53] suggest that the addition of a Graphene coating layer to the Gold metal layer functionalizes the Gold metal layer and due to the carbon based ring structures of Graphene, the absorption of the Gold metal layer increases and thus the sensitivity of the biosensor increases.

Concerning FO-SPR biosensors utilizing a Silver type metal layer, recent studies such as [38] suggest that the addition of a Graphene coat to the Silver metal layer reduces the susceptibility of the Silver to oxidation and corrosion increasing the lifetime of the sensor and functionalizes the Silver to have increased absorption thus increasing the sensitivity of this biosensor.

Our motivation is to decrease the negative effect of the limitations imposed by the Gold and Silver surface plasmon active metal types and due to the extraordinary and advantageous properties of Graphene and due to the multiple recent studies supporting the addition of Graphene as a coating layer, we shall investigate the effect of the addition of a Graphene coating sheet to the Gold and Silver typically used metal layer types for SPR Biosensing in "3.2.3: Simulation of the Effect of the addition of Graphene".

### 3.2.1: Construction and Operation Principle of a Graphene Enhanced and FO-SPR Biosensor

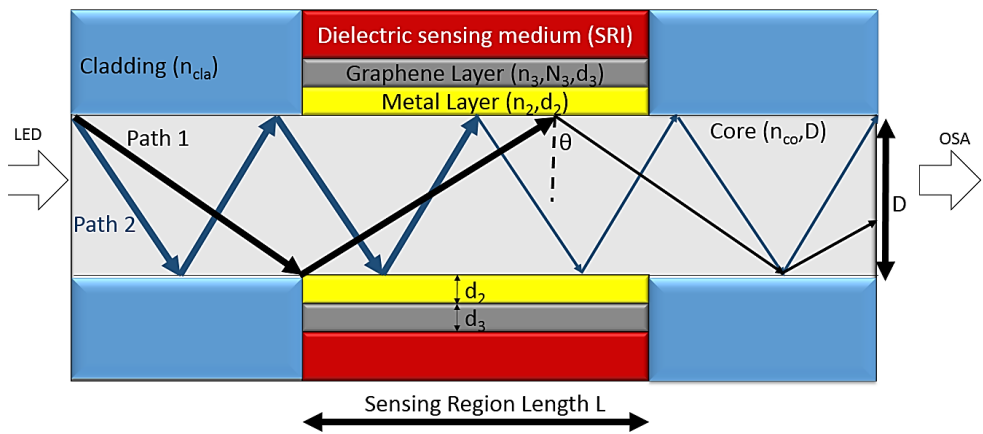

**Figure 16-The construction of a Graphene enhanced and FO-SPR biosensor**

The construction of a Graphene enhanced and FO-SPR biosensor is very similar to the construction of a conventional FO COUPLED SPR biosensor. The only difference is that for the construction of a Graphene enhanced biosensor it is required to add a Graphene coat between the metal and the dielectric sample. So, the sensing region of length $L$ becomes composed of 4 layers:

- The first layer is the MMF core (Diameter $D$ and refractive index $n_{co}$) that only guides light waves of angles of incidence at the core within the acceptance angle of the core and accepted light waves with angles of incidence $\theta$ at the core-cladding interface between the critical angle for TIR at the core-cladding interface $\theta_{cri}$ and 90°. It is important to remind that only P-polarized incident light waves can excite SPWs so the incident light wave should be P-polarized or at least have a P-polarized component.
- The second layer is the metal layer (thickness $d_2$ and dielectric function $\varepsilon_m(\omega)$)that should have a negative real part of the dielectric function to be able to support SPWs at the metal-dielectric interface
- The third layer is the Graphene coat (thickness $d_3$ and refractive index $n_3$ ) used to protect the metal layer against corrosion or oxidation and to functionalize the metal layer surface for increased absorption
- The fourth layer is the dielectric sample layer (refractive index $SRI$) that contains a certain concentration of the analyte of interest that we should find utilizing the biosensor

The operation principle of the Graphene enhanced and FO biosensor is the same as the operation principle of the FO-SPR biosensor without a Graphene coating layer: ATR of the incident light waves at the core-metal layer will cause part of the incident light waves to be absorbed in the form of EWs bound to the core-metal interface and the other part of the incident light waves to be reflected back into the core to be received by the OSA. The component of incident light with a wavelength equal to the resonance wavelength will be maximally absorbed as EWs to excite SPWs at the coated metal-dielectric interface.

The main difference is that the addition of a Graphene layer will alter the reflection coefficient of the reflected light wave off of the sensing region and back into the core causing a variation in the response curve of the biosensor and in the performance parameters of the sensor.

### 3.2.2: Proposed Model for the Analysis of Graphene Enhanced FO-SPR Biosensors

The proposed model for the analysis of the Graphene enhanced and FO-SPR biosensor stratified sensing region is the 4 layer Abélès transfer matrix method proposed in [59] for the 4 layers: fiber optic core, metal layer, Graphene protecting and absorbing layer and the dielectric sample. Each layer is identified by its thickness $d_k$ and dielectric constant $\varepsilon_k$ for $k = 1,2,3,4$.

Since only P-polarized light waves excite SPWs, we have decided to propose an analysis model and simulate the response curves of the sensor assuming that only P-polarized light is incident at the core-metal interface. This could be physically realized by adding a linear polarization lens to polarize the light waves to only have P-polarization.

The refractive index of Graphene in the visible light spectrum is:

$$n_3 = 3 + i\,\frac{5.446 \mu m^{-1}}{3} \lambda \tag{23}$$

And $\lambda$ is the wavelength in vacuum of the incident light wave at the core-metal interface.

The thickness (in nm) of the Graphene layer depends on the number $N_3$ of the 0.34 nm Graphene sheets:

$$d_3 = N_3(0.34) \tag{24}$$

The number of reflections of the incident light waves at the core-metal interface is:

$$N(\theta) = \frac{L}{D \tan(\theta)} \tag{25}$$

And depends on the length of the stratified 4 layer sensing region length $L$, the optical fiber core diameter $D$ and the angle of incidence $\theta$ at the core-metal interface.

The Power of the transmitted P-polarized light wave at the receiving end of the optical fiber is modeled as:

$$P(\theta) = \frac{n_{co}^2 \times sin(\theta) \times cos(\theta)}{(1 - n_{core}^2 \times cos^2(\theta))} \tag{26}$$

And $n_{co}$ is the refraction index of the core of the optical fiber.

The critical angle at which total internal reflection TIR occurs at the core-cladding interface is:

$$\theta_{cr} = sin^{-1}\frac{n_{cla}}{n_{co}} \quad (27)$$

And $n_{cla}$ is the refractive index of the optical fiber cladding.

The normalized transmitted power for the reflected P-polarized light wave off of the core-sensing region interface back into the core is:

$$T = P_{trans} = \frac{\int_{\theta_{cr}}^{\pi/2} R_p^{N(\theta)} P(\theta) d\theta}{\int_{\theta_{cr}}^{\pi/2} P(\theta) d\theta} \quad (28)$$

$R_p$ is the reflection intensity of the one incident light wave at the core-sensing region interface back into the core and is studied utilizing the 4 layer transfer matrix method.

The propagation matrix for each layer $k$ is:

$$M_k = \begin{pmatrix} cos(\beta_k) & \frac{-isin(\beta_k)}{q_k} \\ -iq_k sin(\beta_k) & cos(\beta_k) \end{pmatrix} \quad (29)$$

Where $\beta_k$ is the phase factor for each layer:

$$\beta_k = \frac{2\pi d_k}{\lambda}\sqrt{\varepsilon_k - (n_{co}^2 \times sin^2 \theta)} \quad (30)$$

And where $q_k$ is the optical admittance for each layer:

$$q_k = \frac{\sqrt{\varepsilon_k - (n_{co}^2 \times sin^2(\theta))}}{\varepsilon_k} \quad (31)$$

The 4 layer transfer matrix for the stratified sensing region is:

$$M = \prod_{k=2}^{N-1} M_k = \begin{pmatrix} M_{11} & M_{12} \\ M_{21} & M_{22} \end{pmatrix} \quad (32)$$

And thus the reflection coefficient of the P-polarized light wave at the core-sensing region interface is:

$$r_p = \frac{(M_{11} + M_{12}q_N)q_1 - (M_{21} + M_{22}q_N)}{(M_{11} + M_{12}q_N)q_1 + (M_{21} + M_{22}q_N)} \quad (33)$$

The reflection intensity is thus:

$$R_p = |r_p|^2 \qquad (34)$$

The response curve of the biosensor for spectrum interrogation is the plot of $T$ with respect to $\lambda$.

### 3.2.3: Simulation of the Effect of the addition of Graphene

*Verifying the conditions for SPW excitation and finding the wavelength operation range of the Gold based and Silver Based SPR biosensors:*

Before simulating the response curves of the plasmonic biosensors, we must first confirm that the conditions for excitation of SPWs at the metal-dielectric interface are satisfied as explained in "**Error! Reference source not found.**".

### Condition 1 for excitation of SPWs at the metal-dielectric interface:
- The real part of the dielectric function of the metal should be negative: $Real[\varepsilon_m(\lambda)] < 0$

Since the dielectric function of the metal is dependent on the wavelength of the incident light waves at the core-metal interface, then we should find the range of wavelengths for which the real part of the dielectric function of the metal is negative. This range of wavelengths acts as the range of wavelengths where the metal can support SPW excitation.

As mentioned in "**Error! Reference source not found.**" ,we shall utilize the experimental reflectivity study performed in [10] to find the accurate values of the dielectric function of the noble metals Gold and Silver particularly at the visible and near infra-red frequency ranges that are not supported by the Drude plasma model.

Form the reflectivity study done in [10], we calculate the dielectric function of the Gold type and Silver type noble metal layers form the supplied values of real reflection coefficient $n(\omega)$ and extinction coefficient $\kappa(\omega)$ for a range of incident light photon energies.

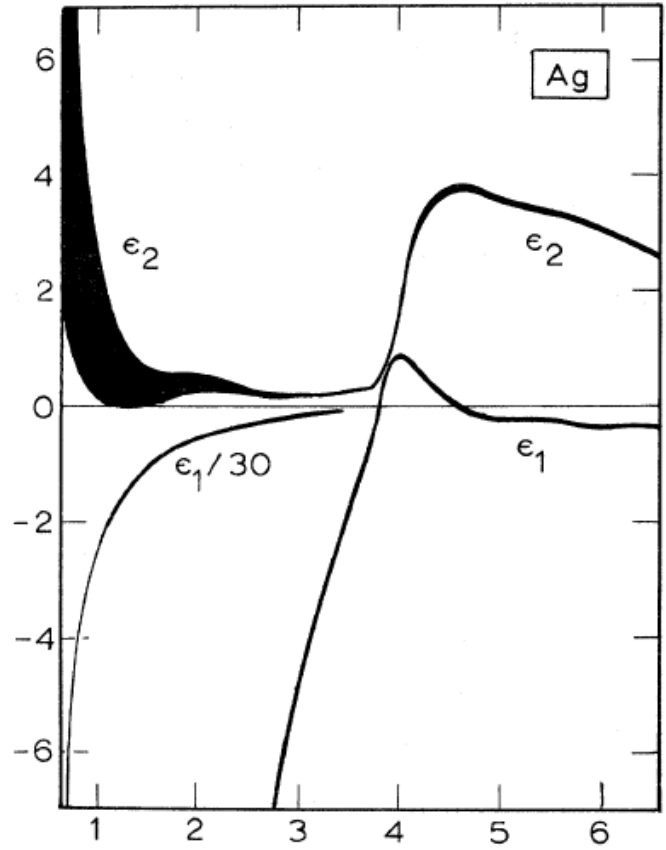

**Figure 17- the dielectrics constants of the Silver metal as a function of Photon energy [10]**

The real part of the dielectric function of the Gold metal is negative for the wavelengths ranging from 292.4 nm to 1393.03 nm hence for SPR biosensors utilizing a Gold type metal layer, the range of operation of the sensor is from 292.4 nm to 1393.03 nm because in this range of wavelengths this sensor metal layer can support SPWs.

| $E_{photon}$ (e.V) | $n(\lambda)$ | $\kappa(\lambda)$ | $\omega$ | $\lambda$ (nm) | Real[ $\varepsilon_m(\lambda)$ ] | Imaginary[ $\varepsilon_m(\lambda)$ ] |
|---|---|---|---|---|---|---|
| 0.89 | 0.43 | 9.519 | 1.35244E+15 | 1393.033708 | -90.426461 | 8.18634 |
| 1.02 | 0.35 | 8.145 | 1.54999E+15 | 1215.490196 | -66.218525 | 5.7015 |
| 1.14 | 0.27 | 7.15 | 1.73234E+15 | 1087.54386 | -51.0496 | 3.861 |
| 1.26 | 0.22 | 6.35 | 1.9147E+15 | 983.968254 | -40.2741 | 2.794 |
| 1.39 | 0.17 | 5.663 | 2.11224E+15 | 891.942446 | -32.040669 | 1.92542 |
| 1.51 | 0.16 | 5.083 | 2.2946E+15 | 821.0596026 | -25.811289 | 1.62656 |
| 1.64 | 0.14 | 4.542 | 2.49214E+15 | 755.9756098 | -20.610164 | 1.27176 |
| 1.76 | 0.13 | 4.103 | 2.6745E+15 | 704.4318182 | -16.817709 | 1.06678 |
| 1.88 | 0.14 | 3.697 | 2.85685E+15 | 659.4680851 | -13.648209 | 1.03516 |
| 2.01 | 0.21 | 3.272 | 3.0544E+15 | 616.8159204 | -10.661884 | 1.37424 |
| 2.13 | 0.29 | 2.863 | 3.23675E+15 | 582.0657277 | -8.112669 | 1.66054 |
| 2.26 | 0.43 | 2.455 | 3.4343E+15 | 548.5840708 | -5.842125 | 2.1113 |
| 2.38 | 0.62 | 2.081 | 3.61665E+15 | 520.9243697 | -3.946161 | 2.58044 |
| 2.5 | 1.04 | 1.833 | 3.799E+15 | 495.92 | -2.278289 | 3.81264 |
| 2.63 | 1.31 | 1.849 | 3.99655E+15 | 471.4068441 | -1.702701 | 4.84438 |
| 2.75 | 1.38 | 1.914 | 4.1789E+15 | 450.8363636 | -1.758996 | 5.28264 |
| 2.88 | 1.45 | 1.948 | 4.37645E+15 | 430.4861111 | -1.692204 | 5.6492 |
| 3 | 1.46 | 1.958 | 4.5588E+15 | 413.2666667 | -1.702164 | 5.71736 |
| 3.12 | 1.47 | 1.952 | 4.74115E+15 | 397.3717949 | -1.649404 | 5.73888 |
| 3.25 | 1.46 | 1.933 | 4.9387E+15 | 381.4769231 | -1.604889 | 5.64436 |
| 3.37 | 1.48 | 1.895 | 5.12105E+15 | 367.8931751 | -1.400625 | 5.6092 |
| 3.5 | 1.5 | 1.866 | 5.3186E+15 | 354.2285714 | -1.231956 | 5.598 |
| 3.62 | 1.48 | 1.871 | 5.50095E+15 | 342.4861878 | -1.310241 | 5.53816 |
| 3.74 | 1.48 | 1.883 | 5.6833E+15 | 331.4973262 | -1.355289 | 5.57368 |
| 3.87 | 1.54 | 1.898 | 5.88085E+15 | 320.3617571 | -1.230804 | 5.84584 |
| 3.99 | 1.53 | 1.893 | 6.0632E+15 | 310.726817 | -1.242549 | 5.79258 |
| 4.12 | 1.53 | 1.889 | 6.26075E+15 | 300.9223301 | -1.227421 | 5.78034 |
| 4.24 | 1.49 | 1.878 | 6.4431E+15 | 292.4056604 | -1.306784 | 5.59644 |

**Table 5- The value of the dielectric function of the Gold metal for a range of incident light photon wavelengths**

The real part of the dielectric function of the Silver metal is negative for the wavelengths ranging from 331.5 nm to 1393.03 nm hence for SPR biosensors utilizing a Silver type metal layer, the range of operation of the sensor is from 331.5 nm to 1393.03 nm because in this range of wavelengths this sensor metal layer can support SPWs.

| $E_{photon}$ (e.V) | $n(\lambda)$ | $\kappa(\lambda)$ | $\omega$ | $\lambda$ (nm) | Real[ $\varepsilon_m(\lambda)$ ] | Imaginary[ $\varepsilon_m(\lambda)$ ] |
|---|---|---|---|---|---|---|
| 0.89 | 0.13 | 10.1 | 1.3524E+15 | 1393.033708 | -101.9931 | 2.626 |
| 1.02 | 0.09 | 8.828 | 1.55E+15 | 1215.490196 | -77.925484 | 1.58904 |
| 1.14 | 0.04 | 7.795 | 1.7323E+15 | 1087.54386 | -60.760425 | 0.6236 |
| 1.26 | 0.04 | 6.992 | 1.9147E+15 | 983.968254 | -48.886464 | 0.55936 |
| 1.39 | 0.04 | 6.312 | 2.1122E+15 | 891.942446 | -39.839744 | 0.50496 |
| 1.51 | 0.04 | 5.727 | 2.2946E+15 | 821.0596026 | -32.796929 | 0.45816 |
| 1.64 | 0.03 | 5.242 | 2.4921E+15 | 755.9756098 | -27.477664 | 0.31452 |
| 1.76 | 0.04 | 4.838 | 2.6745E+15 | 704.4318182 | -23.404644 | 0.38704 |
| 1.88 | 0.05 | 4.483 | 2.8568E+15 | 659.4680851 | -20.094789 | 0.4483 |
| 2.01 | 0.06 | 4.152 | 3.0544E+15 | 616.8159204 | -17.235504 | 0.49824 |
| 2.13 | 0.05 | 3.858 | 3.2367E+15 | 582.0657277 | -14.881664 | 0.3858 |
| 2.26 | 0.06 | 3.586 | 3.4343E+15 | 548.5840708 | -12.855796 | 0.43032 |
| 2.38 | 0.05 | 3.324 | 3.6166E+15 | 520.9243697 | -11.046476 | 0.3324 |
| 2.5 | 0.05 | 3.093 | 3.799E+15 | 495.92 | -9.564149 | 0.3093 |
| 2.63 | 0.05 | 2.869 | 3.9965E+15 | 471.4068441 | -8.228661 | 0.2869 |
| 2.75 | 0.04 | 2.657 | 4.1789E+15 | 450.8363636 | -7.058049 | 0.21256 |
| 2.88 | 0.04 | 2.462 | 4.3764E+15 | 430.4861111 | -6.059844 | 0.19696 |
| 3 | 0.05 | 2.275 | 4.5588E+15 | 413.2666667 | -5.173125 | 0.2275 |
| 3.12 | 0.05 | 2.07 | 4.7412E+15 | 397.3717949 | -4.2824 | 0.207 |
| 3.25 | 0.05 | 1.864 | 4.9387E+15 | 381.4769231 | -3.471996 | 0.1864 |
| 3.37 | 0.07 | 1.657 | 5.1211E+15 | 367.8931751 | -2.740749 | 0.23198 |
| 3.5 | 0.1 | 1.419 | 5.3186E+15 | 354.2285714 | -2.003561 | 0.2838 |
| 3.62 | 0.14 | 1.142 | 5.501E+15 | 342.4861878 | -1.284564 | 0.31976 |
| 3.74 | 0.17 | 0.829 | 5.6833E+15 | 331.4973262 | -0.658341 | 0.28186 |
| 3.87 | 0.81 | 0.392 | 5.8809E+15 | 320.3617571 | 0.502436 | 0.63504 |
| 3.99 | 1.13 | 0.616 | 6.0632E+15 | 310.726817 | 0.897444 | 1.39216 |
| 4.12 | 1.34 | 0.964 | 6.2608E+15 | 300.9223301 | 0.866304 | 2.58352 |
| 4.24 | 1.39 | 1.161 | 6.4431E+15 | 292.4056604 | 0.584179 | 3.22758 |

Table 6- The values of the dielectric function of the Silver metal for a range of incident light photon wavelengths

So in order to satisfy the first condition for excitation of SPWs at the metal-dielectric interface; the poly-chromatic Led light source should have a spectral width within:

- 292.4 nm to 1393.03 nm for SPR sensors utilizing Gold metal layers
- 331.5 nm to 1393.03 nm for SPR sensors utilizing Silver metal layers

We can notice that the operation range of the SPR sensors utilizing Gold or Silver metal layer types includes the visible and infra-red range (near and far) hence the use of the reflectivity study instead of the Drude plasma model was critical for finding an accurate estimation of the refractive index of these metals.

We can also notice that the operation range of the Silver based SPR biosensor is smaller than the operation range of the Gold based SPR biosensor. Hence the Gold metal layer is used as the metal layer of choice of the SPR biosensor when there is an application that has resonance wavelengths out of the operation range of the Silver based sensor but are still within the range of the Gold based sensor despite any performance parameter or adaptability considerations.

Below we represent our curve fitting of the real part of the refractive index $n(\omega)$ and the complex part of the refractive index $\kappa(\omega)$ of the Silver and Gold metal types in order to find the most accurate representation of the dielectric function of these metals.

*Curve fitting for the Gold metal type:*
The typical wavelength range of operation of a Gold based sensor is in the 300 nm to 890 nm wavelength range [29] so, we shall perform curve fitting for this range of wavelengths:

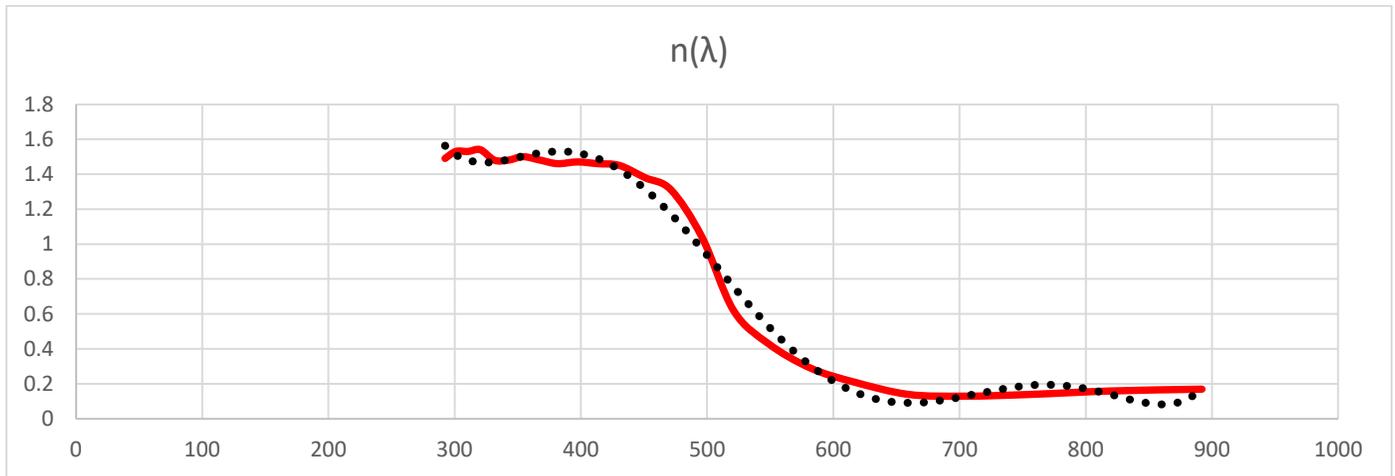

**Figure 18- Curve fitting to find an accurate representation of the real part of the refractive index of the Gold metal type based on the reflectivity study method. The accurate values (red line) are provided by the reflectivity study [10] and the curve fitting values (black dots) form as an approximate of the accurate values.**

From the curve fitting done in "Figure 18" we find the approximate curve fitted equation of the real part of the dielectric function of the Gold metal as a function of wavelength (in nm):

$$n(\lambda) = 0.0000000000000004450099\lambda^6 - 0.00000000001600486251\lambda^5 + 0.000000023237919731228\lambda^4 - 0.0000173551787355187\lambda^3 + 0.00700340457672716\lambda^2 - 1.44744141572266\lambda + 121.452723489391 \quad (35)$$

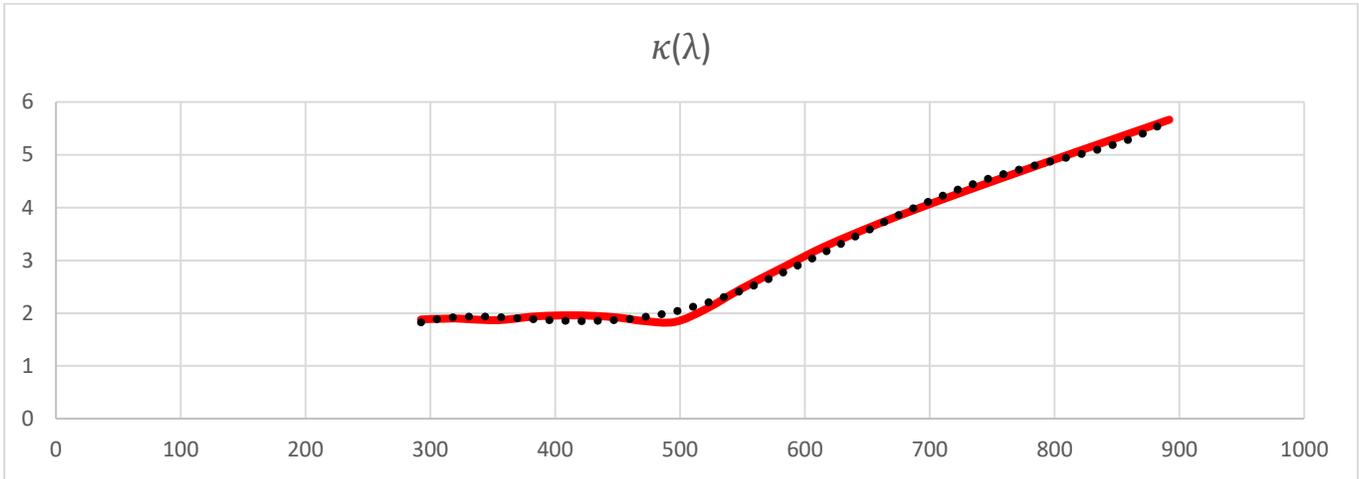

**Figure 19-the accurate values (red line) and curved fitting values (black dots) for the imaginary part of the complex refractive index of the Gold metal type**

From the curve fitting done in "Figure 19" we find the approximate curve fitted equation of the imaginary part of the dielectric function of the Gold metal as a function of wavelength (in nm):

$$\kappa(\lambda)= 0.000000000000000593557\lambda^6 - 0.00000000001230489189\lambda^5 + 0.000000000397530642815\lambda^4 + 0.000000773802078715786\lambda^3 - 0.000734689326188962\lambda^2 + 0.234511353265056\lambda - 23.9207730111468 \quad (36)$$

*Curve fitting for the Silver metal type:*

The typical wavelength range of operation of a Silver based sensor is in the 350 nm to 1215 nm wavelength range [13] so, we shall perform curve fitting for this range of wavelengths:

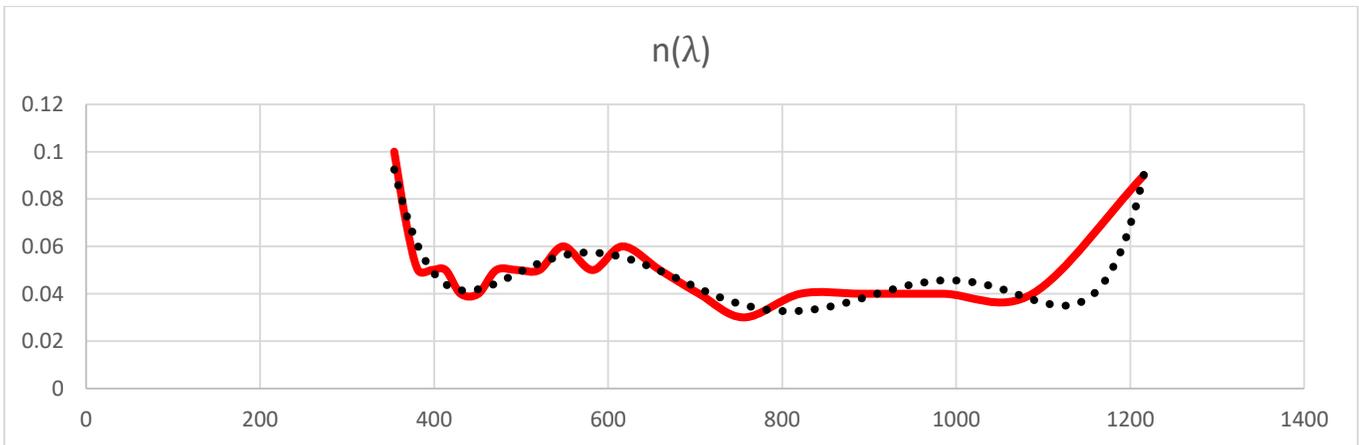

**Figure 20- Curve fitting to find an accurate representation of the real part of the refractive index of the Silver metal type based on the reflectivity study method. The accurate values (red line) are provided by the reflectivity study [10] and the curve fitting values (black dots) form as an approximate of the accurate values.**

From the curve fitting done in "Figure 20" we find the approximate curve fitted equation of the real part of the dielectric function of the Silver metal as a function of wavelength (in nm):

$$n(\lambda) = 0.000000000000000064333\lambda^6 - 0.0000000000000304060006\lambda^5 + 0.000000000583289706258\lambda^4 - 0.00000579813329506956\lambda^3 + 0.000314220010055773\lambda^2 - 0.0878713360762308\lambda + 9.94784820494033 \quad (37)$$

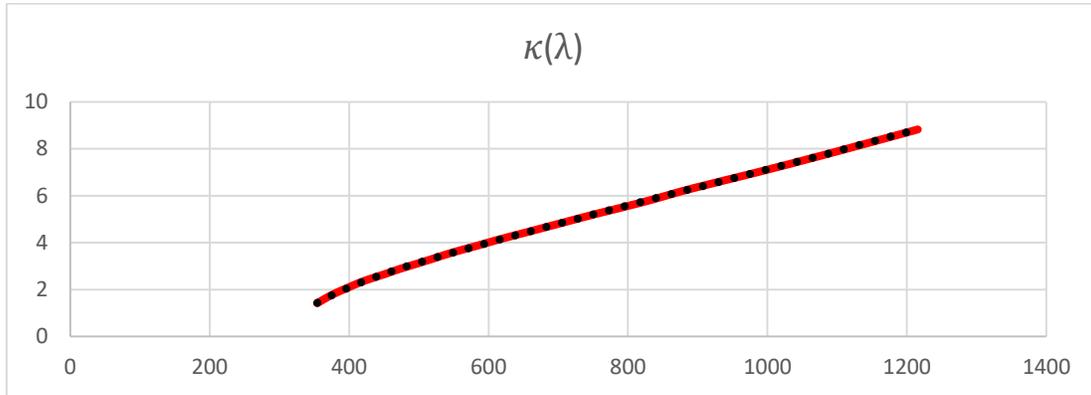

**Figure 21- the accurate values (red line) and curved fitting values (black dots) for the imaginary part of the complex refractive index of the Silver metal type**

From the curve fitting done in "Figure 21" we find the approximate curve fitted equation of the imaginary part of the dielectric function of the Silver metal as a function of wavelength (in nm):

$$\kappa(\lambda) = -0.000000000000000072158\lambda^6 + 0.00000000000365618866\lambda^5 - 0.0000000000758143905657\lambda^4 + 0.00000825141920545449\lambda^3 - 0.000498890275668692\lambda^2 + 0.167459065496102\lambda - 21.9153018564891 \quad (38)$$

We utilize the resulting curve fitted equations in order to find the dielectric function of the Gold and Silver metals according to the formulas and methodology stated in "**Error! Reference source not found.**".

The resulting dielectric function of the Gold and Silver metals should provide an accurate representation of the dielectric constants of these metals for higher photon energies not supported by the Drude plasma model but with errors due to curve fitting.

Condition 2 for excitation of SPWs at the metal-dielectric interface:
- The angular frequency of incident light waves should be less than the plasma frequency: $\omega < \omega_p$

From the effective optical mass and relaxation time values provided in [10], we calculated the values of the plasma frequency for the Gold and Silver metal layer as shown in the table below:

| Metal type | Plasma Frequency $\omega_p$ (rad/s) |
|---|---|
| Gold | 1.792715448190489e+16 |
| Silver | 1.820511143710462e+16 |

Table 7-the Plasma frequency for the Gold and Silver metal types

We can conclude from the wavelength operation ranges of the biosensors that the angular frequencies for the incident light waves within the operation range of the sensor satisfy the second condition for excitation of SPWs.

Condition 3 for excitation of SPWs at the metal-dielectric interface:
- The incident light waves should be P-polarized or at least have a P-polarized component

This condition is satisfied since in the analysis model proposal, we assumed that the incident light waves are only P-polarized.

Optional Condition 4 for excitation of SPWs at the metal-dielectric interface:
- The real part of the dielectric function of the metal has an absolute value that is greater than the imaginary part of the dielectric function of the metal

This is found from "Table 5" and "Table 6" for the Gold and Silver metal layers respectively in their wavelength operation ranges. It is found in the the Silver metal and not found for the Gold metal.

After confirmation that the conditions for the excitation of SPWs are satisfied for a specific range of incident light wavelengths for each metal, we shall simulate the response curves of the FO-SPR biosensor for the Gold and Silver metal layers before and after the addition of the Graphene protecting and absorbing layer within the operation ranges of the SPR sensor for each metal type.

*The response curves of a FO COUPLED SPR biosensor for Gold and Silver metal layers before the addition of Graphene:*

We utilized the values of the dielectric function of the Gold and Silver metal layers from the reflectivity study performed in [10] and we utilized the values of the sensor parameters mentioned in [29] for a conventional FO-SPR biosensor in order to first simulate the response curves for the conventional and still un-optimized FO-SPR biosensor for the Gold and Silver metal layer types respectively. The values of the sensor parameters adopted are stated in "Table 8":

| Parameter | Value |
|---|---|
| Length of the sensing region $L$ | 5 mm |
| Thickness of the metal layer $d_2$ | 40 nm |
| Core refractive index $n_{co}$ | 1.451 |
| Cladding refractive index $n_{cla}$ | 1.450 |
| Core Diameter $D$ | 50 μm |

Table 8- Parameters utilized for the simulation of the response curves of the FO COUPLED SPR biosensor based on the conventional sensor parameters from reference [29]

## 3.3: Simulation for the response curves of the Silver and Gold based biosensors before addition of Graphene:

First we simulate the response curve of the FO-SPR biosensor before the addition of Graphene for varying SRI of the dielectric sample surrounding the Gold and Silver metal layers:

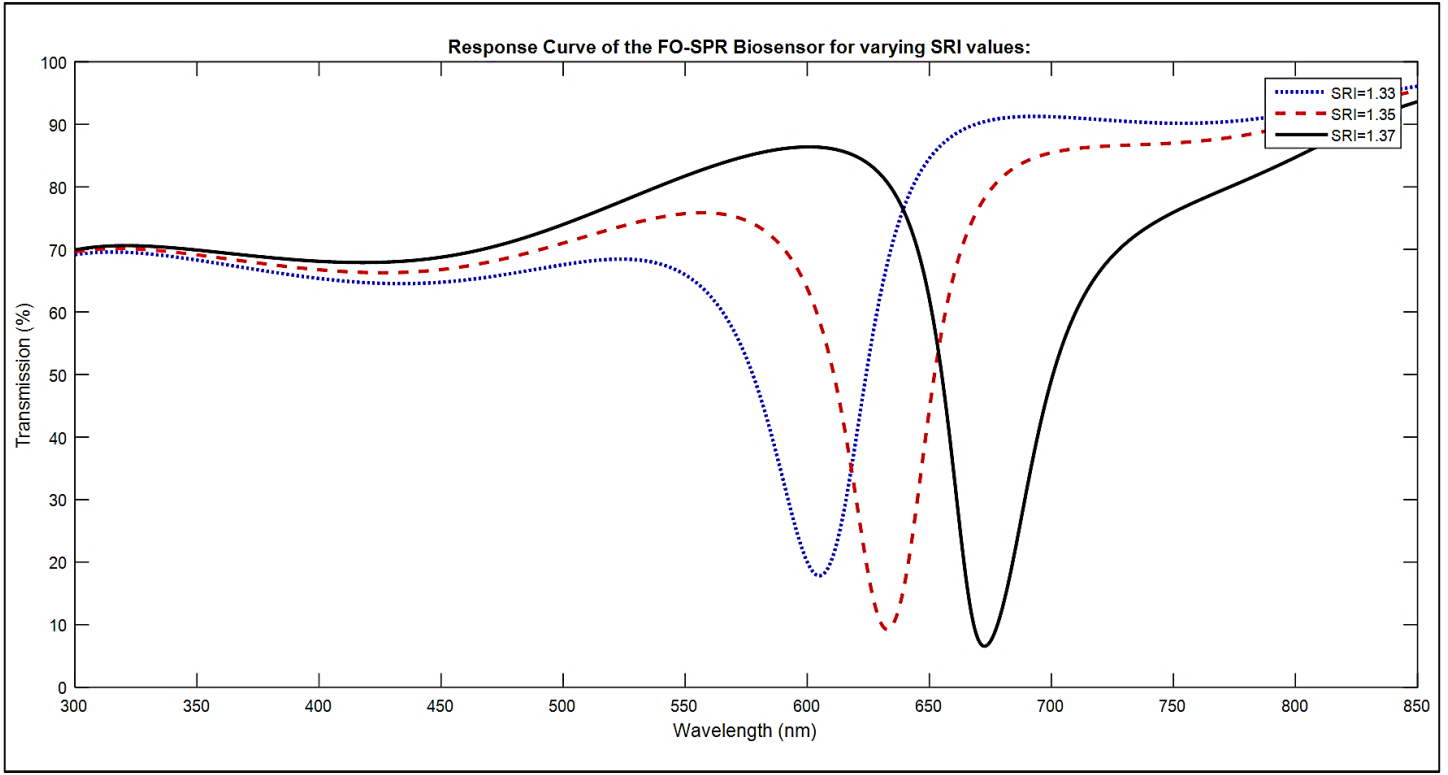

**Figure 22- the response curves of the FO COUPLED SPR biosensor with a Gold metal layer before the addition of Graphene**

| Gold: SRI/data type | Resonance Wavelength (nm) | Minimum of Transmission (%) | Spectral Width at 50% (nm) | Detection Accuracy (TCU%/nm) | Sensitivity with Calibration as Reference (nm/RIU) | Sensitivity with Previous Response Curve as Reference (nm/RIU) |
|---|---|---|---|---|---|---|
| 1.33 | 604.95 | 17.8196 | 46.5 | 1.7673 | Calibration Phase | Calibration Phase |
| 1.35 | 632.8 | 9.2515 | 41.3 | 2.1973 | 1392.5 | 1392.5 |
| 1.37 | 672.6 | 6.5544 | 45.65 | 2.047 | 1691.25 | 1990 |

**Table 9- Table summarizing the values extracted from the response curves of the FO-SPR biosensor with Gold metal layer before the addition of Graphene**

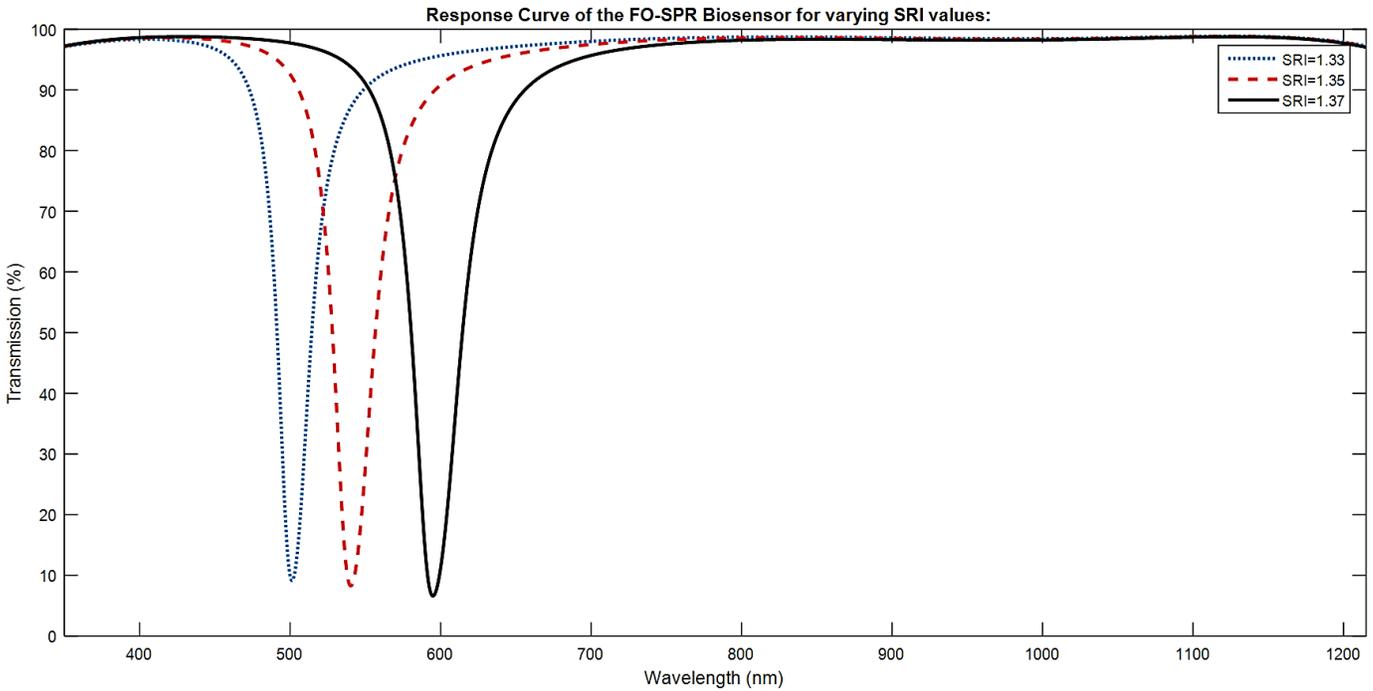

Figure 23- response curves for the FO- SPR biosensor with Silver metal layer before the addition of Graphene

| Silver: SRI/data type | Resonance Wavelength (nm) | Minimum of Transmission (%) | Spectral Width at 50% (nm) | Detection Accuracy (TCU%/nm) | Sensitivity with Calibration as Reference (nm/RIU) | Sensitivity with Previous Response Curve as Reference (nm/RIU) |
|---|---|---|---|---|---|---|
| 1.33 | 501.3 | 9.1257 | 23.05 | 3.9425 | Calibration Phase | Calibration Phase |
| 1.35 | 540.3 | 8.2532 | 28.55 | 3.2135 | 1950 | 1950 |
| 1.37 | 594.95 | 6.5899 | 34.1 | 2.7393 | 2341.25 | 2732.5 |

Table 10- Extracted values from the response curves of the FO-SPR biosensor for Silver metal type before the addition of Graphene

We can conclude the following from the response curves of the biosensor for varying SRI before the addition of Graphene while utilizing either the Gold or the Silver layer as a general conclusion for the output seen at the OSA for any SPR biosensor design:

- As the SRI of the dielectric sample surrounding the metal increases, there will be a red-shift in the resonance wavelength towards the higher wavelengths hence in simulations we must insure that the resonant wavelength of our particular application stays within the frequency operation range of the sensor which could dictate the type of metal layer to be used
- As the SRI increases, the minimum of transmission at the resonance wavelength decreases hence the dip of transmission at the resonance wavelength increases in depth as the SRI increases
- As the SRI increases, the spectral width of the response curve at 50% transmission increases hence there will be broadening of the response curve as SRI increases

- As the SRI increases, the sensitivity of the response curve increases because the increase in the SRI causes a red-shift in resonance wavelengths
- As the SRI increases , there is a tradeoff between obtaining a deeper dip but for the negative effect for having a broader response curve spectral width

When comparing the response curves of the FO- SPR biosensors with same sensor parameters but for differing metal layer type between Gold and Silver, we can conclude:

- The Silver based sensor exhibits a smaller spectral width at 50% transmission compared to the Gold based sensor (about half of the initial spectral width of the Gold based sensor at the calibration phase) hence the Silver based sensor has a narrower response curve.
- The Silver based sensor exhibits a smaller minimum in transmission at the resonance wavelength compared to the Gold based sensor (about half the value of the minimum of transmission of the Gold based sensor at the calibration phase) hence for the Silver based sensor there is a greater dip depth in transmission at the resonance wavelength.
- The silver based sensor has a greater detection accuracy than the Gold based sensor
- The Silver based sensor has a greater sensitivity than the Gold based sensor

In both the Silver based sensor and Gold based sensor, there is a broadening in the spectral width of the response curves as the SRI increases but for Silver based sensors this broadening has a more limited negative effect on the detection accuracy of the sensor because its initial spectral width is considerably smaller compared to that of the Gold based sensor.

The response curve dip at the resonance wavelength increases in depth as the SRI increases for both the Silver based sensor and the Gold based sensor so, the issue of having a shallow dip at the resonance wavelength in the calibration phase response curve is not as critical and comes secondary to the issue of having a very broad spectral width at the calibration phase response curve because the spectral width only increases as SRI increase thus, the spectral width is the main parameter that compromises the detection accuracy of the sensor.

From the discussion above we can see that the Silver based SPR biosensor provides higher detection accuracy and higher sensitivity when compared to the Gold based sensor. The only issue with utilizing Silver as the metal layer of the SPR biosensor is the fact that Silver is not immune to oxidation and corrosion like Gold. Addition of Graphene to Silver has been proved to reduce the oxidation and corrosion of the Silver metal layer [13] and thus increase the lifetime of Silver based sensors so, now we should investigate the effect of the addition of Graphene to both the Silver based sensor and Gold based sensors to see the affect that Graphene will induce on the response curves of the sensors.

### 3.4: Simulations of the response curves of the biosensor after the addition of Graphene:

We now investigate the effect that the addition of a Graphene layer has on the response curves of the SPR biosensor as the SRI increases for both the Gold and Silver metal layer types.

We shall utilize the same sensor parameters stated in reference [29] and summarized in "Table 8" for a conventional un-optimized FO-SPR biosensor. The same sensor parameters that were utilized to generate sensor response curves before the addition of Graphene are utilized to generate the response curves of the sensor after the addition of Graphene in order to confirm that the only variable affecting the variation in the response curves of the sensor is the addition of a Graphene layer and not any other factor.

We shall study the effect of adding 5 Graphene sheets of 0.34 nm thickness each to the Gold based sensor and Silver based sensor:

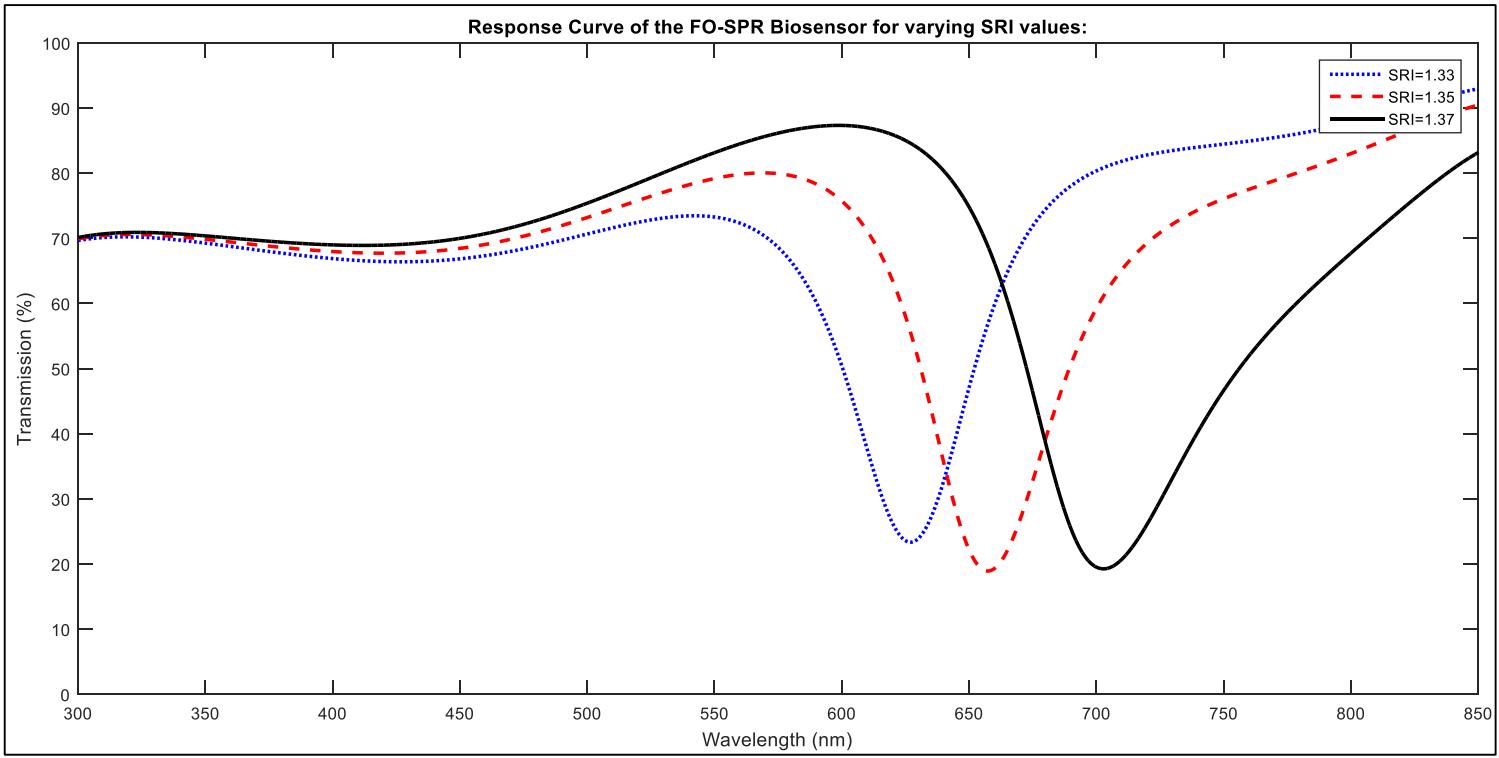

Figure 24-the response curve of the Gold based sensor for the addition of 5 layers of Graphene

| Gold: SRI/data type | Resonance Wavelength (nm) | Minimum of Transmission (%) | Spectral Width at 50% (nm) | Detection Accuracy (TCU%/nm) | Sensitivity with Calibration as Reference (nm/RIU) | Sensitivity with Previous Response Curve as Reference (nm/RIU) |
|---|---|---|---|---|---|---|
| 1.33 | 627 | 23.3549 | 51.75 | 1.4811 | Calibration Phase | Calibration Phase |
| 1.35 | 657.35 | 18.931 | 58.4 | 1.3882 | 1517.5 | 1517.5 |
| 1.37 | 702.8 | 19.2636 | 83.5 | 0.9669 | 1895 | 2272.5 |

44Table 11-the values extracted from the response curves of the Gold based SPR biosensor for the addition of 5 sheets of Graphene

| Silver: SRI/data type | Resonance Wavelength (nm) | Minimum of Transmission (%) | Spectral Width at 50% (nm) | Detection Accuracy (TCU%/nm) | Sensitivity with Calibration as Reference (nm/RIU) | Sensitivity with Previous Response Curve as Reference (nm/RIU) |
|---|---|---|---|---|---|---|
| 1.33 | 536.9 | 30.8893 | 38.95 | 1.7743 | Calibration Phase | Calibration Phase |
| 1.35 | 579.7 | 26.9777 | 48.8 | 1.4964 | 2140 | 2140 |
| 1.37 | 638.75 | 22.1067 | 60.9 | 1.279 | 2546.25 | 2952.5 |

Table 12-the values extracted from the response curves of the Silver based SPR biosensor for the addition of 5 Graphene sheets

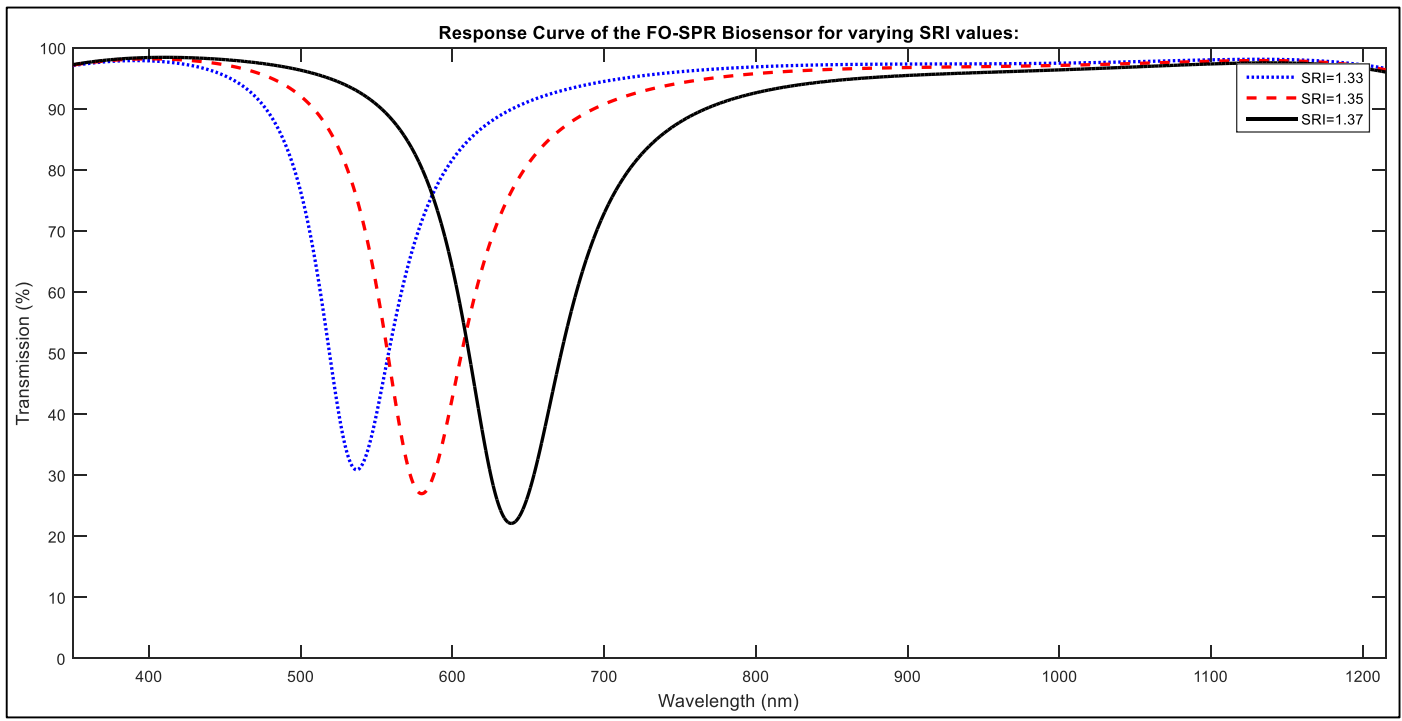

Figure 25-the response curves of the Silver based SPR biosensor for the addition of 5 sheets of Graphene

We can conclude the general effect that the addition of Graphene has on the response curves of the SPR biosensors and applicable for the Silver based and Gold based SPR biosensors:

- As the SRI increases, the behavior of the response curve of the biosensor before and after the addition of Graphene is the same. Hence when the SRI increases, the spectral width of the response curve becomes broader, the resonance wavelength experiences a red-shift and the dip in transmission at the resonance wavelength becomes deeper
- The addition of Graphene causes a red-shift in the resonance wavelengths towards the higher wavelengths compared to the resonance wavelengths of the same biosensor but without the addition of a Graphene layer. Hence upon choosing to add a Graphene layer to the sensor, it should be confirmed that the resonance wavelengths do not exceed the operation frequency range of the sensor

- The addition of Graphene causes the minimum of transmission at the resonance wavelength to be higher that then minimum of transition for the same biosensor but without the addition of Graphene. Hence, the addition of Graphene causes the response curves to become shallower compared to the response curves of the biosensor before the addition of Graphene
- The addition of Graphene causes the spectral widths of the response curves to be greater than the spectral widths of the response curves of the sensor before the addition of Graphene. Hence the addition of Graphene led to the broadening of the response curves compared to the response curves of the sensor before the addition of Graphene
- The sensitivity of the graphene enhanced biosensor is greater than the sensitivity of the same sensor without the addition of Graphene

We can conclude from the general effect that the addition of a Graphene layer has on the response curves of the biosensor; that for the addition of Graphene there is a tradeoff between increased sensitivity and decreased dip and decreased width selectivity so decreased detection accuracy.

We should also compare between the response curves of the Graphene enhanced biosensors with Gold and Silver as the metal layers:

- After the addition of Graphene, the sensitivity of the Silver based sensor is becomes higher than the sensitivity of the Silver based sensor before the addition of Graphene and even higher than the sensitivity of the Graphene enhanced Gold based sensor
- After the addition of Graphene to the Silver based sensor, the response curves of the Graphene enhanced Silver based sensor become broader and shallower compared to the Silver based sensor before the addition of Graphene. Even after the addition of Graphene to the Silver based sensor, the detection accuracy of the sensor remains higher than or comparable to the detection accuracy of the Gold based sensor even before the addition of Graphene

## 3.5: Conclusions to be implemented for optimization:

From the analysis done above, we decide to select the metal layer type of the biosensor as Silver since the Silver based sensor provides the higher sensitivity and the higher detection accuracy compared to the Gold based sensor for before and after the addition of Graphene.

We also decide that the addition of Graphene should be implemented since it increases the lifetime of the Silver based sensor by protecting against corrosion and also because Graphene increases the absorption and hence sensitivity of the Silver based sensor noticeably and does not have a great negative impact on the detection accuracy of the Silver based sensor since the detection accuracy of the Graphene enhanced Silver based sensor is comparable to the detection accuracy of the Gold based SPR biosensor before the addition of Graphene.

46Conclusion:

In our project we extensively discussed the theory behind designing a plasmonic biosensor that is: nano-sized, label-free, optical, implantable and disposable and provides real-time continuous monitoring of the concentration of the desired analyte.

We also simulated the response curves of the biosensor in order to find an optimized design for this biosensor that respects both sensitivity and detection accuracy considerations as well as practicality for in situ biomedical sensing applications.

In our study, we discussed an emerging recent nanotechnology which is Plasmonics nanotechnology and adopted a subset of Plasmonics which is surface plasmon wave generation and utilized the concept surface plasmon wave generation to design a label-free, real-time and nano-sized optical biosensor that eliminates the possibility of electrocution.

In our study, we also tackled an emerging search field which is the improvement of sensing devices by introduction Graphene and concluded that the addition of Graphene did in fact cause a notable increase in sensitivity of the sensor without increasing the complexity of the design and without compromising the nano-size and biocompatibility of the sensor. Depending on our simulations, we concluded that the only drawback for the addition of a Graphene layer to the biosensor was the decrease in detection accuracy of this biosensor but since the pros of the addition of Graphene outweighed the cons, we decided to add Graphene as a protective and absorptive coat for the metal layer of the sensor.

We optimized the parameters of the sensor by studying the effect that each parameter had on the response curves and on the performance parameters of the sensor and selected the values of the parameters that guarantee satisfactory sensor performance.

We also adapted the optimized and Graphene enhanced plasmonic SPR biosensor to be disposable and implantable by comparing between the conventional T-type design and the R-type design. After comparing the simulations we performed for each type of sensor, we concluded that despite the R-type sensor decreasing the sensitivity negligibly and the detection accuracy modestly we should still adopt the R-type design because it is very advantageous when it comes to practicality of disposal of the sensor probe and in implanting the sensor probe in the human body.

After deciding on the finalized design of the biosensor, we gave an example of how this general design of the biosensor could be adaptable to measure Glucose concentration in a sample of water depending on the Glucose oxidase bio-receptor that alters the SRI of the water sample surrounding the sensor probe depending exclusively on the variation in the concentration of Glucose.

We proposed instrumentation needed to physically implement the sensor in order to estimate an approximate cost of the biosensing system to give an in-depth analysis of the safety considerations.

Limitations of the proposed Biosensor:

In our biosensing system design, we combated the aging of the Silver metal layer expressed in the form of cracking by the addition of a Graphene coat but we did not add a mechanism for combating the aging of the LED light source because we want to be as cost-effective as possible and we assumed that the monitoring session duration would not be long enough that the LED light source would vary significantly in its performance.

Our Publications and Conferences:

Concerning the topic of our final year project, we presented one conference oral presentation and published three publications to IEEE xplore:

- One conference attended which is the "1st student innovation and research conference for Arab universities and schools SIRCAUS" where we submitted an accepted oral presentation about the effects of Graphene on the enhancement of the sensitivity of the SPR biosensor
- One Publication titled "On the Design of Graphene Surface Plasmon Resonance Sensors for Medical Applications" in the '2018 IEEE International Symposium on Antennas and Propagation and USNC-URSI Radio Science Meeting' held in Massachusetts, Boston on the 19th to the 22nd of July 2018
- Two publications titled "Overview of Fiber Optic Surface Plasmon Resonance Biosensors for Medical Applications" and "On the Enhancement of Surface Plasmon Resonance Biosensors by Graphene for Medical Applications" in the ANTEM 2018 conference in Waterloo in the special session title "Applied Electromagnetic" and I am expected to present the two accepted publications on August 19th to the 22nd in Waterloo, Canada.

Our Future prospects:

In the near future we intend be the first authors to physically implement the novel design of the optimized T-type and Silver based FO-SPR biosensor with and without the Graphene layer to compare between the theoretical and practical results obtained before and after the addition of Graphene and to confirm the advantages of the addition of Graphene and also to particularly test the effect that Graphene has on the lifetime of the Silver based sensor.

Later we intend to fabricate an R-type alternative for the biosensor and compare it with the conventional T-type design in vivo. Then we intend to further study the bio-receptor part of the sensor and to try to physically realize the immobilization of artificial bio-receptors on the surface of the metal probe with the help of professionals in the fields of biology.

We are also interested in studying further configuration alternatives that provide immunity from the degradation of biosensor performance due to aging of the biosensing system.

The projected topic of our next publication is expected to be about the comparison between the Gold and Silver type based sensors for the addition of Graphene and optimization of the parameters of the sensor that has a Silver metal layer which is the metal layer that provided the best performance parameters.